\documentclass[apj]{emulateapj}
\usepackage{subfigure}
\usepackage{amsmath}
\usepackage{amssymb}
\usepackage{bm}
\usepackage{url}
\newcommand{\mdeg}{\ensuremath{^{\circ}}}
\usepackage[usenames,dvipsnames]{xcolor}
\submitted{}
\journalinfo{Accepted for publication in The Astrophysical Journal}
\begin{document}
%%%
%%%
\title{Inter-relationship between the Two Emission Cones of B1237$+$25}
\author{ Yogesh Maan\altaffilmark{$\dagger$} and Avinash A. Deshpande}
\affil{Raman Research Institute, Bangalore -- 560080, India}
\email{yogesh@rri.res.in \& desh@rri.res.in}
\altaffiltext{$\dagger$}{Joint Astronomy Programme (JAP), Indian Institute of Science, Bangalore -- 560012, India}
\begin{abstract}
The origin of two distinct pairs of conal emission components in pulsars,
associated with the ``outer'' and the ``inner'' emission cones, as well
as the marked difference in their observed spectral properties, is poorly
understood.
The sub-pulse modulation in the corresponding conal components, if mapped
back to the underlying system of sub-beams rotating around the magnetic
axis in the polar cap, as envisioned by Ruderman~\&~Sutherland~(1975),
provides a potential way to investigate the emission morphologies in the
two conal regions, and more importantly, any inter-relationship between them.
The bright pulsar B1237+25 with its special viewing geometry where the
sightline traverses almost through the magnetic axis, along with a rich
variety in pulse-to-pulse fluctuations, provides an excellent, but
challenging opportunity to map the underlying emission patterns across
the full transverse slice of its polar emission region. We present here
our analysis on a number of pulse-sequences from this star to map and
study any relationship between the underlying patterns responsible for
emission in the two pairs of presumed conal-components and a core-component
of this pulsar. The results from our correlation analysis of the two
conal emission patterns strongly support the view that the two cones
of this pulsar (the outer and the inner cone) originate from a common
system of sub-beams. We also see evidence for a twist in the emission
columns, most likely associated with a corresponding twist in the
magnetic field structure. We discuss these results, and their implications,
including a possibility that the core component of this pulsar
shares its origin partly with the conal counterparts.
\end{abstract}
%%=======================================================================
\section{Introduction}
The phenomenon of ``sub-pulse drifting'', i.e., the systematic
variation in position and intensity of sub-pulses, was noticed \citep{DC68}
soon after the discovery of pulsars.
Ruderman \& Sutherland (1975; hereafter R\&S) suggested this regular
modulation to be a manifestation of a carousel of ``spark'' discharges
circulating around the magnetic axis in the acceleration zone of the
star because of the E$\times$B drift. After almost two decades of the
above proposition, \citet{DR99,DR01} developed a cartographic transform
to map the systematic sub-pulse variations to a carousel of sub-beams
projected on to the polar cap. They traced back the origin of the coherent
modulation in pulse sequences of B0943+10 to a system of 20 sub-beams
circulating around the magnetic axis. Their detailed study provided
strong support to the carousel model proposed by R\&S. The cartographically
mapped emission pattern represents a slice of the emission cone at an
altitude corresponding to the frequency of observation \citep[as suggested
by the radius-to-frequency mapping (RFM);][]{Cordes78}. The deduced
emission-map can provide valuable details of the radio emission (for
example, spatial structure and distribution of sub-beams, their
temporal and spectral evolution, etc.) in the sampled region of
the polar magnetosphere.
\par
In the above picture of a carousel of emission columns, the viewing
geometry limits the accessible region of the slice in the polar
magnetosphere which can be mapped. A tangential sightline traverse across
the sub-beam ring(s) allows us to sample only the periphery of the
carousel pattern, and the circulation of sub-beams would manifest
as the apparent sub-pulse ``drifting'' (i.e., primarily phase modulation
within the pulse-profile). In case the sightline-traverse is near/through
the magnetic pole (i.e., impact angle $\ll$ cone radius), the modulation
of sub-pulses will appear primarily as an amplitude modulation, and one
would be able to sample the emission pattern, and hence the cone(s),
more completely. The well-known bright pulsar B1237$+$25 offers one such
unique example of viewing geometry. The multi-component average profile
of this pulsar consists of two pairs of conal components and a core
component. Such a profile is presumably produced by a
sightline which samples two concentric emission cones and a central core
beam almost through their common center. Using a sequence of pulses where
the core-emission was almost absent, \citet[see their Figure 8; hereafter
SR05]{SR05} have shown a full ``conal'' position-angle traverse with an
exceptionally steep sweep rate at the center
($-185\mdeg\pm5\mdeg\,{\rm deg}^{-1}$, may be the largest ever observed),
as expected from the magnetic pole model \citep{RC69}. Their estimate
for the impact angle ($\beta\lesssim 0.25\mdeg$) confirms that our
sightline indeed traverses very near (or probably through) the
magnetic axis of this pulsar.
\par
As expected from the viewing geometry of this pulsar, the prominent type of
observed sub-pulse modulation is indeed the amplitude modulation of the pairs
of conal-components. The core-component
consists of a complex structure \citep{SRM13}, and
has been reported to be ``incomplete'' (SR05).
Unlike many other
pulsars, where the fluctuation spectrum of the core-component is generally
featureless \citep{Rankin86}, the core component of this pulsar shows presence
of some features at very low fluctuation frequencies (see, e.g., SR05).
B1237$+$25 also exhibits two different emission modes --- ``normal''
and ``abnormal'' modes \citep{Backer70a,Backer70b,Backer70c,Backer73}
--- distinguished by significant changes in conal as well as core
emission, total power and polarization properties. The normal mode
exhibits a regular $2.7$-period sub-pulse modulation in its conal
components, while this modulation ceases in the abnormal mode.
The normal mode
is further divided into ``quiet-normal'' mode and ``flare-normal'' mode,
to separate pulses with weak core-activity from those with active or
`flared-up' core emission (SR05). These two (sub-)modes
were also found to have different polarization properties. Low
fluctuation frequency features exhibited by the core component are
believed to originate from the quasi-periodic interruption of the
quiet-normal mode pulse sequences by the flare-normal mode instances.
\par
Like several other multi-component (``M''-category) pulsars, B1237$+$25
also exhibits emission components corresponding to both the cones ---
the ``inner'' as well as the ``outer'' cone.
However, the co-existence of emission from both the cones
in M-stars, and possible inter-relationship between them, if any, have
remained poorly understood. An intriguing possibility that `the inner cone
is emitted at a lower height along the same group of peripheral field lines
that produce the outer cone' had been suggested a long time ago by
\citet{Rankin93a}. In the case of pulsar B0329$+$54, \citet{GG01} show
that the multiple cones appear to originate at different heights in the
magnetosphere but along relatively nearby field lines. This supports the
above possibility suggested by \citet{Rankin93a}, and has remained the only,
but an indirect, evidence so far. The polar emission patterns corresponding
to the two cones, if could be mapped, can help in finding out whether the
emission in the two cones really share a common origin or not. These maps
can also be useful in studying any other relationship between the inner
and outer emission cones.
\par
With a rich variety in its pulse-to-pulse fluctuations and the sightline
traverse being very close to the magnetic axis, B1237$+$25 provides an
important and challenging opportunity to study the characteristics of
emission across the entire transverse extent of its polar emission cone,
and specially till
close vicinity of the magnetic axis. With the specific aim of studying
any inter-relationship between the sub-beam patterns responsible for
emission in the two cones, we have analyzed four different
pulse sequences (we denote them as A, B, C and D, comprising
5209, 2340, 5094 and 4542 pulses, respectively) from this
star observed at 327 MHz using the Arecibo telescope\footnote{These
327-MHz pulse sequences (A, B, C and D) were acquired using the Wideband
Arecibo Pulsar Processor (WAPP2 ) on 2005 Jan 09, 2003 July 13, 14 and 21,
respectively. The effects of dispersion and interstellar Faraday rotation
across the observation bandwidth were corrected for, and various
instrumental polarization effects were removed before producing the pulse
sequences. Pulse sequence A has a resolution equivalent to 0\mdeg.133
in pulse longitude, while that in the other sequences is 0\mdeg.352.
Further details on the observations of pulse sequence A, and sequences
B,C and D can be found in \citet{SRM13} and \citet{SR05},
respectively.}.
In this paper, we present our analysis of these
pulse-sequences\footnote{We are thankful to Joanna Rankin for
making these pulse sequences and the polarization as well as emission
mode separated versions of
sequence A \citep[reported in detail by][]{SRM13} available to us.},
including the clarifying mode-separated versions of sequence A,
to map the sub-pulse fluctuations to the underlying patterns of
circulating sub-beams.
The sub-sequences used for constructing the emission maps
primarily consist of ``normal'' mode sequences with minor contamination
from ``abnormal'' mode, as detailed later.
The reconstructed emission maps were then subjected to correlation
analysis to study any interrelated properties of the emission patterns
associated with the two conal rings and the central core-beam.
The preliminary results of this work were reported earlier in the conference
``40 Years of Pulsars : Millisecond Pulsars, Magnetars and More'' \citep{MD08}.
\par
In section~\ref{sect_analysis}, we describe the analysis procedures we have
followed to deduce and verify the sub-beam circulation period, and present
polar emission maps corresponding to the pulse sequences. A possible
inconsistency between the polar emission maps and the standard carousel
model, in terms
of the carousel rotation phase, is discussed in section~\ref{sect_modphase}.
Details of the correlation analysis of the emission maps are provided in
section~\ref{sect_cones}, followed by a discussion on the results and their
implications in section~\ref{sect_discussion}. Conclusions drawn from
our study are presented in section~\ref{sect_conclusions}.
%%=======================================================================
%%=======================================================================
%%=======================================================================
\section{The polar magnetosphere emission patterns}\label{sect_analysis}
%%%
The pulse-to-pulse fluctuations in the above mentioned four single pulse
sequences (A, B, C and D) were studied in detail, particularly to
examine if a
carousel of sub-beams can explain the rich modulations. Below we present
our analysis procedure to examine the fluctuation properties over a range of
timescales to estimate and verify the carousel circulation period, 
followed by reconstructed emission maps using a number of selected
sub-sequences.
%%%
\subsection{Fluctuation spectral analysis}
To map the systematic sub-pulse fluctuations to a rotating
carousel of sub-beams in the polar magnetosphere, we need
to know the carousel circulation period ($P_4$, i.e., the tertiary
modulation period) and the viewing geometry of the pulsar (i.e.,
the magnetic inclination angle, $\alpha$, and the sightline impact
angle, $\beta$).
To find out the circulation period, we examine the fluctuation power
spectrum, as well as the auto/cross-correlation function, of the
longitude-resolved intensity sequences. In the domain of fluctuation
spectrum, the circulation period may reveal itself by the presence of
one or both of the following features:
\begin{enumerate}
\item A low-frequency feature which directly corresponds to the
circulation period,
\item A signature of amplitude modulation, due to the sub-beam intensity
pattern in the presumed carousel, in the form of a pair of symmetric
side-bands about the secondary\footnote{Various
modulations in a typical intensity time-sequence are generally termed
as follows: primary modulation refers to the modulation due to
the pulsar rotation itself, the secondary modulation corresponds to the
sub-pulse modulation (i.e., ``drifting'' or amplitude modulation), and
tertiary modulation is that due to rotation of the carousel.} modulation
feature (corresponding to $P_3$).
\end{enumerate}
The information content in the auto/cross-correlation function is in
principle the same
as in the fluctuation power spectrum. However, signatures of any tertiary
modulation may be more prominently seen in the correlation function when the
secondary fluctuation features are of very low quality\footnote{Generally
assessed by estimating the $Q$-factor of the corresponding spectral feature
($Q=\frac{f}{\Delta f}$).} or if the circulation period itself is not
constant. Particularly, a cross-correlation between fluctuations of the
two pulse components which are expected to be associated with the same
conal-ring (and hence sharing the same underlying modulation, but with
a relative delay) can be very revealing. In any case, to establish the
relevance of a spectral feature to the carousel circulation period, it
is necessary to confirm the consistency of any delay expected between
the fluctuations of such pulse components with that apparent from either
the cross-correlation function or from the phase-gradient in the
corresponding cross-spectrum.
%%%
\par
The longitude-resolved fluctuation spectrum (LRFS) of this pulsar
(see, for example, figure 6 of SR05) consists of a broad feature
at around 0.35 cycles/Period (hereafter c/P, with the rotation
period $P\approx1.382$ s). The low Q-value of this
secondary modulation feature makes it difficult to identify the presence
of any side-bands associated with the tertiary modulation. The large width
of this feature along with the absence of any obvious low-frequency feature
(corresponding to the carousel circulation) in the LRFS suggests that the
underlying pattern of sub-beams is not stable over long durations
($\sim1000\,P$), i.e., on the timescales over which the fluctuation
properties are being examined. The irregularity may lie within the
underlying emission pattern, i.e., in the inter-spacings and intensities
of the sub-beams, as well as in the circulation period of the carousel.
%%%=========================================================================
\begin{figure}
\centering
 \includegraphics[width=0.5\textwidth,angle=-90]{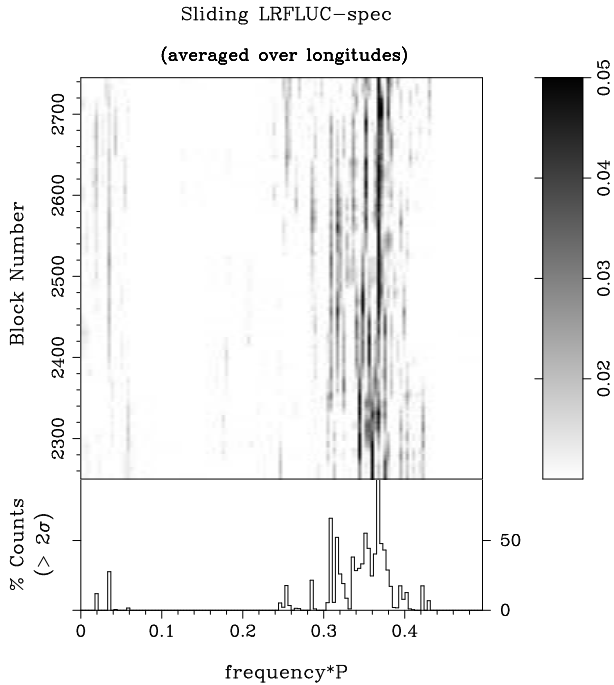}
 \caption{\textsl{Sliding Fluctuation Spectrum:} Each row in the main
panel shows the fluctuation spectrum of a sub-sequence whose starting pulse
number is marked as `Block Number' in the left panel. The bottom panel shows
the percentage number of times the spectral power in each of the frequency
bins crossed a threshold of $2\,\sigma$.}
 \label{sfs_a1}
\end{figure}
%%%-------------------------------------------------------------------------
%%%=========================================================================
\par
To assess stability of the presumed carousel on shorter timescales,
we make use of the ``sliding'' fluctuation spectrum
\citep[SFS; introduced by] [and called S2DFS]{SSW09}\footnote{The
only difference in SFS and S2DFS is in the procedure used to obtain
the individual spectra in the stack. In S2DFS, 2-dimensional fluctuation
spectrum is averaged to compute each of the spectra in the stack,
while SFS makes use of the LRFS for the same, as mentioned in the
text.}. SFS is a stack of fluctuation spectra computed using an
n-period wide window which slides by a pre-decided number of pulses
each time. Each of the fluctuation spectra in the stack is obtained
by first computing the LRFS for the corresponding window, and then
averaging over all the longitudes. Using the SFS, we can explore
the stability and time-evolution of a modulation feature, as well
as detect spectral features corresponding to relatively short-interval
tertiary modulations. Figure~\ref{sfs_a1} shows an example of SFS computed
for a sequence comprising of 750 pulses, with a 256-period wide window
slid across the pulse sequence. Each time, the fluctuation power
spectrum integrated over all the longitudes was computed, and the window
was slid forward by one pulse. Corresponding to each of the fluctuation
spectra (rows in the main panel of Figure~\ref{sfs_a1}), the starting
pulse number is marked as `Block Number' in the left panel. The bottom
panel shows the percentage of time when the spectral-power crosses a
chosen threshold at each of the frequency bins. This panel is specifically
helpful in detecting a modulation feature associated with circulation
of pattern(s) lacking stability over long durations, but appearing
intermittently with an otherwise stable period. A low frequency feature
$\sim0.035$ c/P apparent in Figure~\ref{sfs_a1}, is seen to be prominent
only between the Block Numbers about 2450--2600.
%%%
\par
For each of the four pulse sequences, we examined two sets of SFS, computed
using window widths of 256 and 512 periods. These sets were examined
specifically for possible presence of any high-Q secondary modulation
feature, side-bands around it or a low frequency feature which may directly
relate to the circulation period. Sub-sequences of appropriate lengths,
corresponding to instances of significant spectral features in the SFS,
were selected and investigated further for assessing credibility of the
candidate circulation period suggested by the relevant features.
%%%
%%%-------------------------------------------------------------------------
\begin{figure*}
\centering
\subfigure[]{\includegraphics[width=0.5\textwidth,angle=-90]{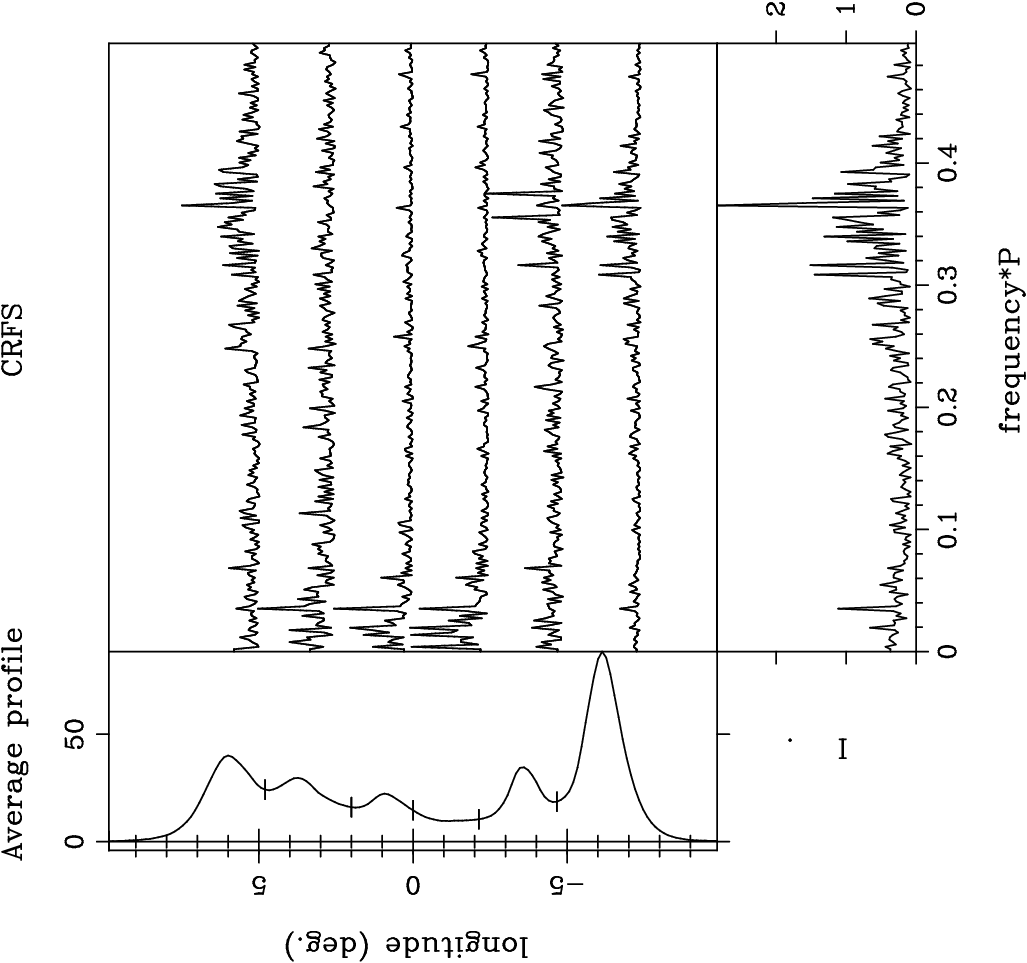} \label{fluc_summary_a1}}
\subfigure[]{\includegraphics[width=0.5\textwidth,angle=-90]{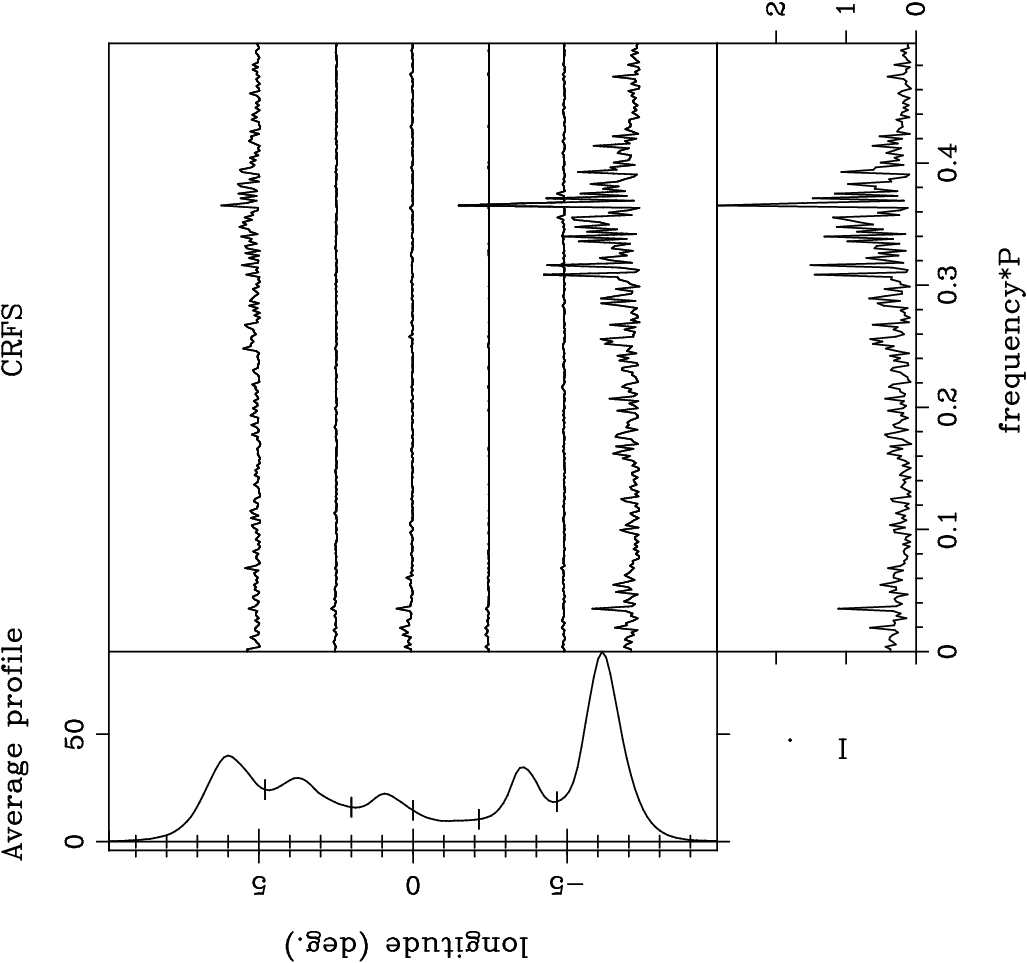} \label{fluc_summary_a2}}
\caption{\textsl{Normalized and Unnormalized Component-Resolved
Fluctuation Spectra:} In each of the two sub-figures, the main panel shows
the average fluctuation spectra corresponding to the sections marked in
the average pulse-profile plotted in the left panel, computed for the
sub-sequence A$_1$. To present the details in these fluctuation spectra
more clearly, sub-figure~(a) shows the spectra normalized by the respective
peak values. Sub-figure~(b) presents the spectra without any normalization,
to enable comparison of modulation power corresponding to various
frequency features in different components.
The bottom panel (in both the sub-figures) shows the integrated spectrum
(i.e., the average of fluctuation spectra corresponding to all the
longitudes within the profile).}
\label{fluc_summary_a}
\end{figure*}
%%%-------------------------------------------------------------------------
%%%
\subsection{The emission maps}\label{section_maps}
After the above described analysis of the pulse sequences A, B, C \& D,
four sub-sequences (hereafter denoted by A$_1$, B$_1$, C$_1$ \& D$_1$,
each a subset of the pulse sequence suggested by it's name) which showed
one or more of the desired spectral signatures were chosen for further
investigations.
The pulse-number ranges that define the sub-sequences A$_1$,
B$_1$, C$_1$ and D$_1$ within the respective parent sequences are:
2391--2990, 1529--1784, 3410--3921 and 2685--3196, respectively.
To verify the candidate circulation period, for each of
these sub-sequences, the longitude-longitude correlation maps \citep[][]{PS90}
were examined to check if the fluctuations at pulse longitudes associated
with the same emission cone (i.e., the same carousel) show the expected
phase relationship (or lag). The auto/cross-correlation functions of
intensity sequences corresponding to selected components were also examined
for consistency of the observed phase relationship with the expected
modulation. After such verification, the estimated circulation period
was further refined by using the `closure' path provided by the inverse
cartographic transform described in \citet{DR01}. In this technique,
best-$P_4$ is found out by searching in a small range around the
candidate $P_4$, and using cross-correlation (between the original
pulse sequence and that synthesized from an emission map constructed
using a given trial $P_4$ value) as figure of merit \citep{Desh00}.
%%%
%%%
\par
\begin{figure}
\centering
 \includegraphics[width=0.45\textwidth,angle=-90]{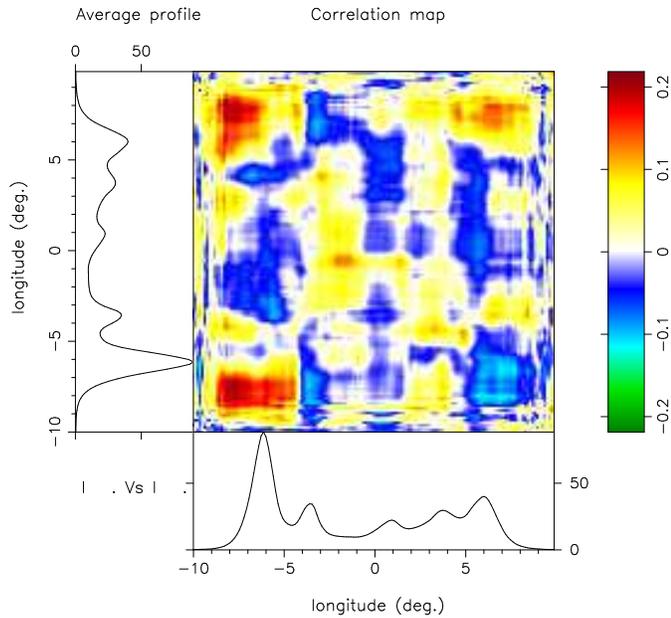}
\caption{\textsl{Longitude-Longitude Correlation Map:} The central
panel shows the cross-correlation of intensity fluctuations at
different longitudes across the pulse window at a relative delay of 14
pulse periods, for the sub-sequence A$_1$. The bottom and left panels show
the average profiles corresponding to the original and delayed sub-sequence,
respectively. A significant correlation between the first and last component
is evident, consistent with a circulation period of about 28.5 P.}
 \label{fluc_summary_b}
\end{figure}

Figure~\ref{fluc_summary_a}~\&~\ref{fluc_summary_b} summarize the
fluctuation properties, and
present evidence for the tertiary modulation in the sub-sequence
A$_1$ (this is the longest among the selected four sub-sequences).
Figure~\ref{fluc_summary_a1} presents the fluctuation spectrum averaged
separately over longitude sections identified with
each of the components. The respective sections are as marked in the
average pulse-profile plotted on the left hand side. Each of these spectra
are normalized to their respective peak values. Merely to differentiate
it from the conventional LRFS, we refer to it as component-resolved
fluctuation spectrum (CRFS). A low-frequency feature at
$0.0351\pm0.0002$ c/P is visible across all the sections, suggesting a
circulation period of $28.5\pm0.2$ P.
Note that the largest fraction of the modulation power corresponding
to this low frequency feature is in the first component, and not in
the core, as evident from the \emph{unnormalized} CRFS shown in
Figure~\ref{fluc_summary_a2}.
Further, since our line of sight
cuts almost through the magnetic pole of the pulsar, one would expect
a phase difference of about 180$^\circ$ between the intensity fluctuations
in the pulse components corresponding to the same emission cone. This is
consistent with the presence of significant correlation between the first
and last components when correlated after a delay of 14 P (as apparent from
the longitude-longitude correlation map
shown in Figure~\ref{fluc_summary_b}). The correlation
between the inner conal components is also visible, although it is not
so prominent. These evidences support the view that the low frequency
feature in the fluctuation spectrum is due to the circulation periodicity
we are seeking. A feature around $0.0196\pm0.0004$ c/P lies close
to the first sub-harmonic of above feature, and has contribution
primarily from the central region of the profile, i.e., from the
core-component and the region prior to it.
%%%
%%%-------------------------------------------------------------------------
%%%-------------------------------------------------------------------------
\par
\begin{figure}
\centering
 \includegraphics[width=0.45\textwidth,angle=-90]{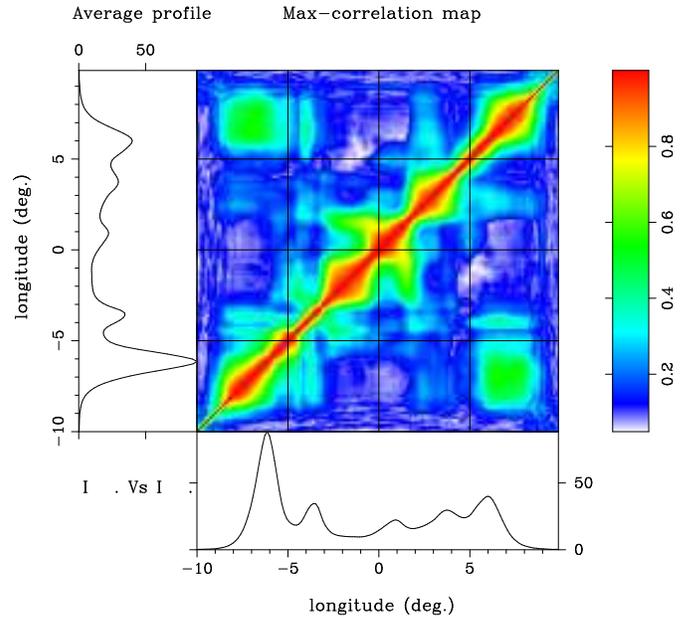}
\caption{\textsl{Longitude-Longitude Max-correlation Map:} The central
panel shows the \textit{maximum} cross-correlation coefficient of intensity
fluctuations at different longitudes across the pulse window within a
relative delay range of $-30$ to $+30$ pulse periods, for the sub-sequence
A$_1$. The bottom and left panels show the average profiles corresponding to the original and delayed sub-sequence,
respectively. No significant correlation is seen between the core intensity
fluctuations and those in the conal components. The significant correlation
between the two outer conal components is due to the secondary modulation,
as suggested by the corresponding delays.}
 \label{maxcorr_map}
\end{figure}

Note that the quasi-periodic flare-normal mode apparitions manifest
a slow amplitude modulation in the central part of the profile,
resulting in low frequency feature(s) in the fluctuation spectrum.
This modulation has been suggested to affect a broad region around
the core-component, including the inner conal components (SR05).
Although the origin of the semi-periodic appearances of the flare-normal
mode pulse sequences is not clear, a recent study reports very different
dynamics of the core radiation \citep{SRM13}, suggesting that the core
component might not have its origin in a carousel-like system.
Hence, it is important to assess whether the above $\sim28.5$ P
modulation observed in the conal components also has its origin
in the slow amplitude modulation of the core-region. To investigate
any correlated intensity fluctuations between the core and conal
components, and more generally between any two longitudes of the
profile, correlation function was computed for each pair of longitudes
for our sub-sequence A$_1$.
The maximum correlation coefficient at each of the longitude-pairs
(in the delay range $-30$ to $+30$ P, more than appropriate for the
observed modulation of $\sim28.5$ P) was then plotted similar to that
in longitude-longitude correlation map. The results of this correlation
analysis convincingly show that the overall modulations in the core-region
has no significant correlation with those in the conal components
(see Figure~\ref{maxcorr_map}; although there are hints of weak correlation
with those in the inter-conal region).
%%%-------------------------------------------------------------------------
%%%-------------------------------------------------------------------------
%%%-------------------------------------------------------------------------
\begin{figure}
\centering
 \includegraphics[width=0.3\textwidth,angle=-90]{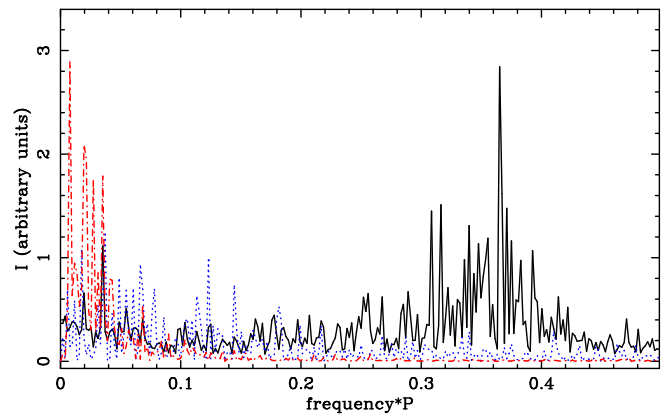}
 \caption{\textsl{Comparison of various spectra:}
The spectra of the sequences of nulls and flare-normal mode pulses 
shown by the dotted (blue) and dash-dotted (red) lines, respectively,
are overlaid on the integrated fluctuation spectrum of the sub-sequence
A$_1$ (the continuous black line), for a ready comparison. The total
power in the former two spectra are normalized to the power in the
lower frequency half of the later spectrum, for ease of comparison.}
 \label{composite_spec}
\end{figure}
%%%-------------------------------------------------------------------------
\par
We note that the core component exhibits several low frequency features,
and the feature under consideration contributes $<10\%$ of the total
spectral power of this component. Our correlation analysis described
above does not exclusively probe the possible correlation contribution 
corresponding to only this particular low frequency feature. 
To further explore potential contribution of the flare-normal mode
and null apparitions towards the low frequency modulation, we separately
examined the fluctuation spectra of pulse sequences associated with
the flare-normal 
mode and nulls, respectively\footnote{For this purpose, we constructed
sequences with a one-to-one mapping between the sequence-elements and the
pulse numbers in our sub-sequence A$_1$. For constructing the sequence
associated with the flare-normal mode, the sequence element is assigned a
value of 1 if the corresponding pulse in the sub-sequence is identified
as a flare-normal mode pulse, and 0 otherwise. The spectrum of this
sequence of 1s and 0s was then examined. A sequence of 1s and 0s was
constructed for nulls also in a similar manner, and the corresponding
spectrum was examined.}.
The fluctuation spectra of the null and flare-normal mode sequences,
overlaid on the integrated spectrum of the sub-sequence, are shown in
Figure~\ref{composite_spec}. Note that the spectra associated with
nulls and flare-normal mode do exhibit a range of features, including
those corresponding to the two low frequency modulation features in
the integrated spectrum ($\sim$0.019~c/P and 0.035~c/P). Clearly,
there are also several other low frequency features, which do not have
any correspondence in the integrated spectrum, making it highly unlikely
for the two low frequency modulation features to be selectively 
originated due to nulls/flare-normal mode\footnote{We have also
examined two \emph{longitude resolved cross-spectra},
computed by using the core and conal fluctuation spectra as references.
Both the cross-spectra show the two relevant low frequency features
throughout the profile. The cross-spectrum computed using the core
spectrum as a reference shows additional prominent low frequency
features \emph{confined} to the core region of the profile, most
likely caused by the quasi-periodic apparitions of the flare-normal
mode. Presence of these additional features only in the core region
again makes it highly unlikely that the two low frequency modulation
features would have selectively originated due to flare-normal mode.}.
Additionally, it is unexpected for the two unrelated phenomena, i.e.,
the nulls and the flare-normal mode, to exhibit common intrinsic
modulations. Presence of such common modulation features therefore
suggests a common external cause, such as carousel circulation.
Hence, the above comparative assessment of various fluctuation spectra
reassures
that the observed conal modulation of $\sim28.5$ P does not have its
origin in the quasi-periodic apparitions of the flare-normal mode or
nulls. In fact, this observed conal modulation resulting from carousel
circulation adds corresponding features to the spectra of the
nulls/flare-normal mode.
%%%
%%%-------------------------------------------------------------------------
\begin{figure}
\centering
\includegraphics[width=0.45\textwidth,angle=-90]{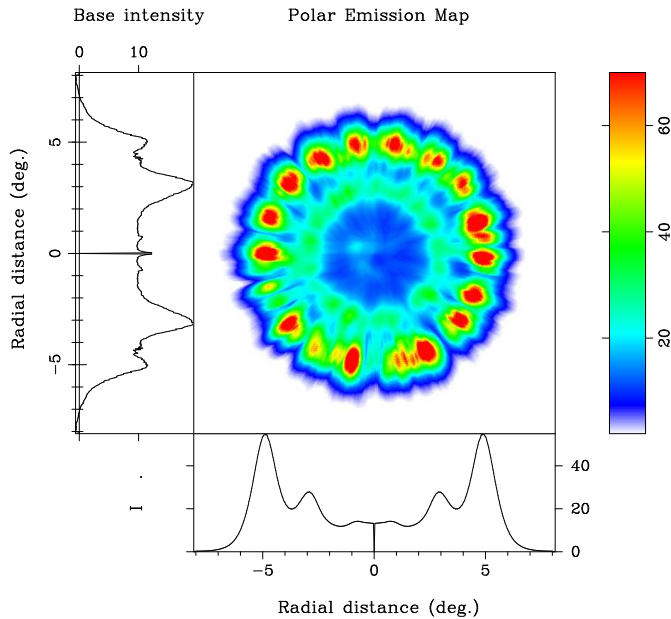}
\caption{\textsl{Polar Emission Map:} An image of the accessible emission
region of the pulsar B1237$+$25 at 327 MHz, constructed using the sub-sequence
A$_1$, with the geometrical parameters and circulation period value mentioned
in the text. The map of the emission
region, shown in the main panel, is projected on to the polar cap. The bottom
and the left-hand side panels show the average- and the base-intensity
profiles, respectively, as functions of the angular distance from the
magnetic axis. Note the easily distinguishable sub-beams in the outer and
the inner conal-rings at radial distances of about $5$ and $3$ degrees,
respectively.}
\label{polar_map}
\end{figure}
%%%-------------------------------------------------------------------------
\par
A map\footnote{The mid-point between the pair of conal components, which
was found to be same for both the pairs within our resolution, is taken as
the ``zero''-longitude.}
of the polar emission region constructed using a further refined
value of the circulation period ($28.41\,P$, by applying the `closure'
path mentioned earlier), and the geometrical
parameters ($\alpha=53.0$$^\circ$, $\beta=0.25$$^\circ$)
from SR05, is shown in Figure~\ref{polar_map}. Presence of 18
bright sub-beams in the outer emission cone is easily noticeable. Some
of the sub-beams appear to be bifurcated or corrugated, however that
is not unexpected, given the discrete spread seen in the secondary feature.
Interestingly, the secondary modulation period
$P_3$, calculated by dividing $P_4$ by the number of sub-beams,
corresponds to the aliased value of the secondary modulation feature.
This indicates that the observed feature around $\sim0.37\;$c/P might
actually be a first order alias of its actual value around $\sim0.63\;$c/P.
%%%
\par
A few percent ($\approx5\%$) of the pulses in our sub-sequence A$_1$
show characteristics of the abnormal mode (as judged from
mode-separated sequences). However, by analyzing this
sub-sequence excluding the abnormal mode pulses, we have confirmed
that the inclusion of these few percent pulses do not have any
noticeable effect (neither qualitatively nor quantitatively) on
the reconstructed emission map, as well as on the results of the
subsequent correlation analysis presented in Section 4.
%%%
%%%-------------------------------------------------------------------------
\begin{figure*}
\centering
\subfigure[]{\includegraphics[scale=0.44,angle=-90]{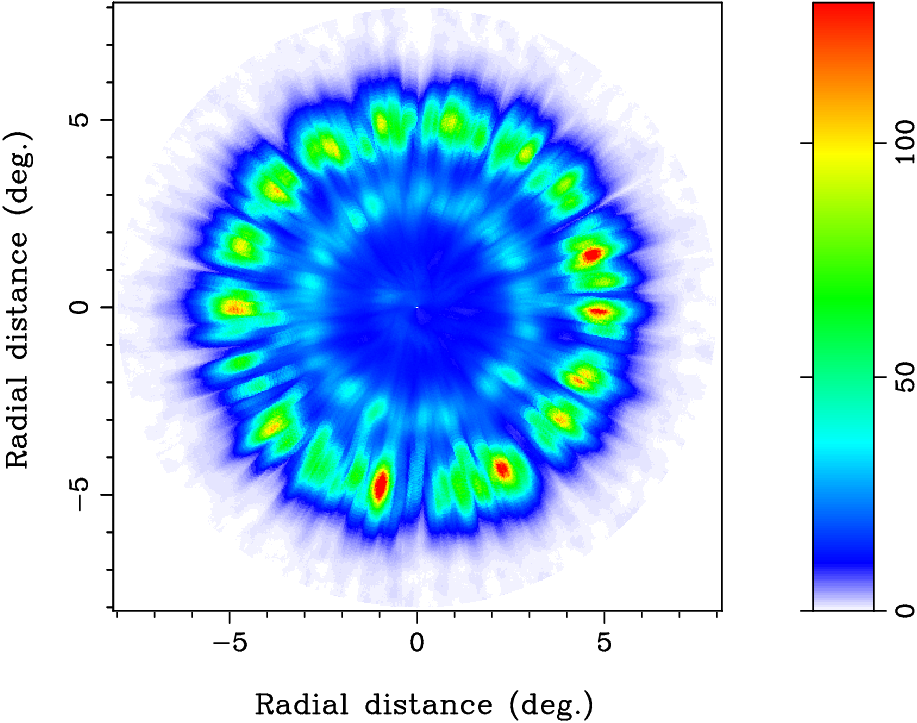}
 \label{polar_maps_all_a}}
\hspace{0.5cm}
\subfigure[]{\includegraphics[scale=0.44,angle=-90]{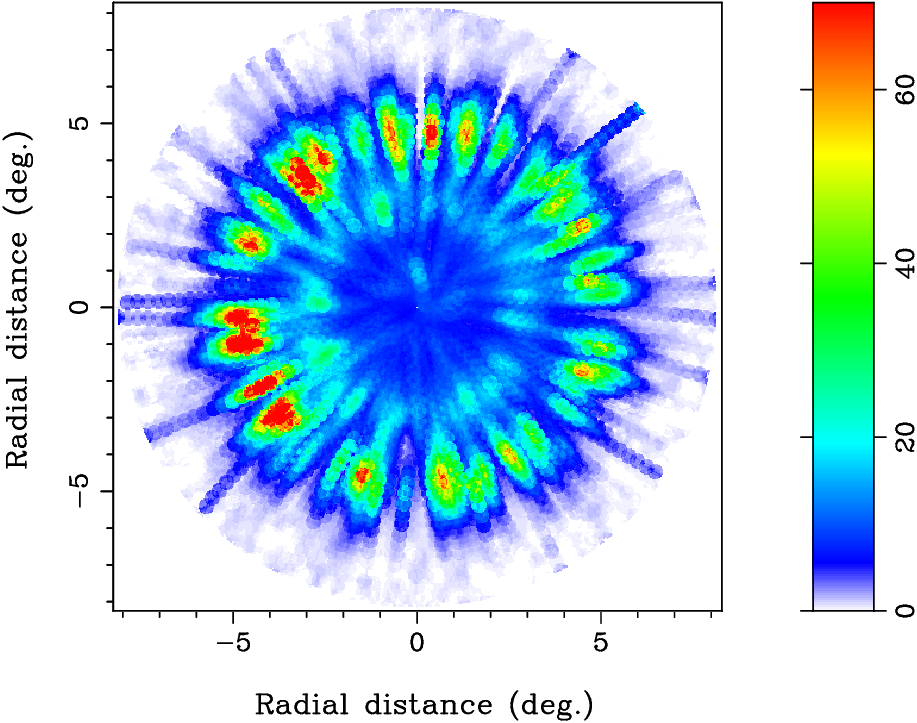}}
\subfigure[]{\includegraphics[scale=0.44,angle=-90]{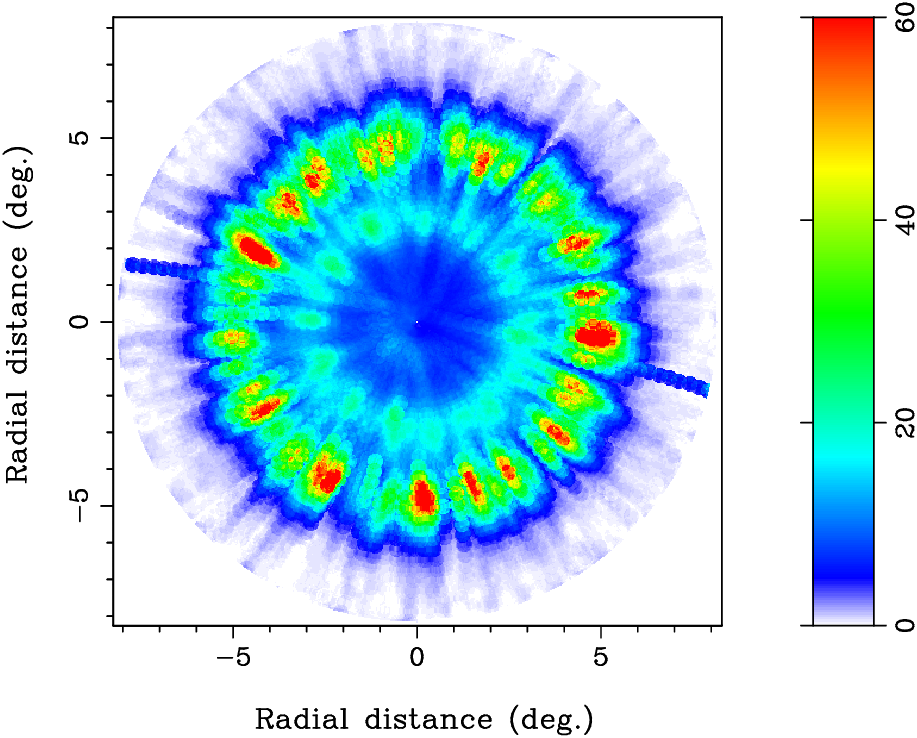}}
\hspace{0.5cm}
\subfigure[]{\includegraphics[scale=0.44,angle=-90]{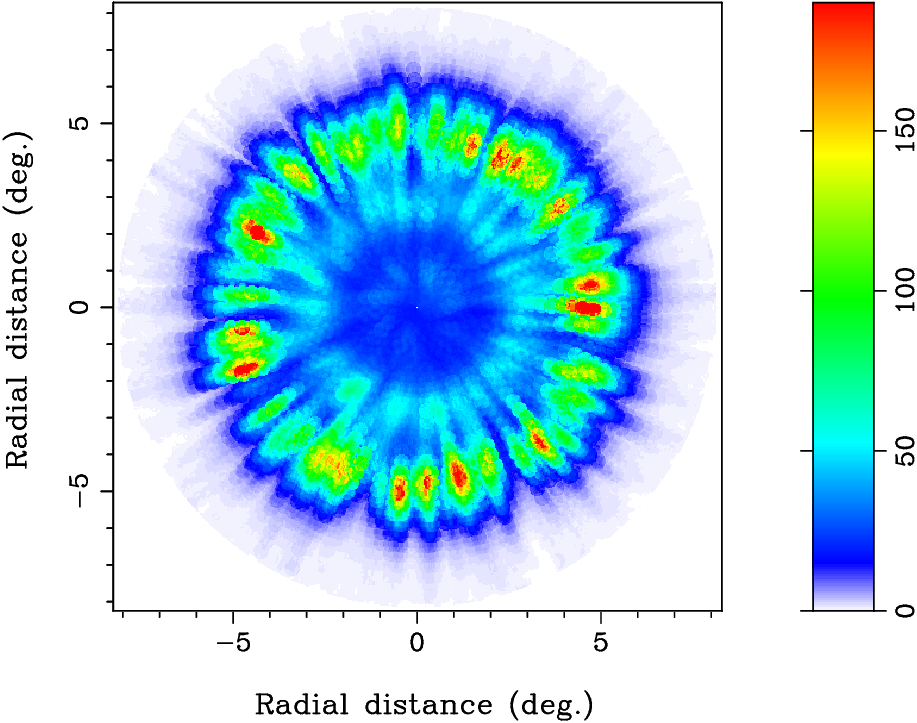}}
\caption{Polar emission maps constructed for various sub-sequences
(A$_1$: Top left; B$_1$: Top right; C$_1$: Bottom left; \&
D$_1$: Bottom right) using the corresponding circulation periods
mentioned in Table~\ref{mod_summary}. These maps have been smoothed
minimally, to show the details at small spatial scales.}
\label{polar_maps_all}
\end{figure*}
%%%-------------------------------------------------------------------------
%%%=========================================================================
\begin{deluxetable*}{cccccccc}
\tabletypesize{\footnotesize}
\tablecolumns{8}
\tablewidth{0pt}
\tablecaption{Modulation parameter summary of various sub-sequences.}
\tablehead{\colhead{Sub-seq.}                                           &
	   \colhead{No. of}						&
           \colhead{$P_{\rm LF}$($P$)}                			&
	   \colhead{$P_{\rm HF}$($P$)}					&
	   \multicolumn{2}{c}{No. of sub-beams\tablenotemark{$\dagger$}}	&
	   \colhead{Deduced\tablenotemark{$\ddagger$}}			&
           \colhead{Null}                                  		\\
	   \cline{5-6} \noalign{\smallskip}
	   \colhead{}							&
	   \colhead{Pulses}						&
	   \colhead{}							&
	   \colhead{}							&
	   \colhead{Outer}						&
	   \colhead{Inner}						&
	   \colhead{$P_3$($P$)}						&
	   \colhead{Fraction(\%)}					\\
	   \colhead{}							&
	   \colhead{}							&
	   \colhead{}							&
	   \colhead{}							&
	   \colhead{ring}						&
	   \colhead{ring}						&
	   \colhead{}							&
	   \colhead{}						
	  }
\startdata
%---------------------------------------------------------------------------
A$_1$ &600 &$28.41\pm0.16$ &2.5--3.0 &18   &18   & 1.58~[2.72]  &05.0$\pm$0.9 \\%2.841
%---------------------------------------------------------------------------
B$_1$ &256 &$18.27\pm0.15$ &2.5--3.1 &12   &4/12 & 1.52~[2.92]  &11.3$\pm$2.1 \\%2.837
%---------------------------------------------------------------------------
C$_1$ &512 &$33.88\pm0.27$ &2.6--2.9 &13/22 &12  & 1.54~[2.85]  &04.7$\pm$1.0 \\%2.822
%---------------------------------------------------------------------------
D$_1$ &512 &$23.23\pm0.19$ &2.6--3.1 &8 &6--8/15 & 1.55~[2.82]  &05.5$\pm$1.0 \\%2.873
%---------------------------------------------------------------------------
\enddata
\tablecomments{(1) $P_{\rm LF}$ is the period deduced using the observed
\emph{low fluctuation frequency} feature. $P_{\rm LF}$ is treated as $P_4$
in our analysis, and values refined using the closure path, as
described in the text, are presented here. The corresponding uncertainties
are estimated using the fluctuation spectrum.
(2) $P_{\rm HF}$, i.e., period deduced using high frequency feature, most
likely corresponds to $P_3$, and the range corresponding
to the discrete spread seen in the secondary feature is presented here.
(3) For sub-sequences B$_1$, C$_1$ and D$_1$, the sub-beam spacing in
the outer ring, and hence the corresponding number of sub-beams, could not
be estimated unambiguously.}
\tablenotetext{$\dagger$}{Note that when brightness of individual sub-beams
is highly non-uniform (e.g., when only a few of them are bright), the number
of sub-beams are prone to be underestimated. Also, varying sub-beam spacing,
and incomplete sampling of the emission map (especially for a sub-sequence
of small length, e.g., B$_1$, and towards the larger magnetic colatitudes)
further limit the certainty with which the number of sub-beams can be
estimated. Due to these reasons, the estimates of no. of sub-beams are
specifically less reliable for the last three sub-sequences, i.e., for B$_1$,
C$_1$ and D$_1$.}
\tablenotetext{$\ddagger$}{$P_{\rm LF}$ divided by the estimated number of
sub-beams provides the ``deduced $P_3$'' estimates. In case of ambiguity
in the number of sub-beams, the largest value for the number of sub-beams
is used. The square brackets present the periods that LRFS would indicate
(after first order aliasing) for the corresponding ''deduced $P_3$'' values.}
\label{mod_summary}
\end{deluxetable*}
%%%=========================================================================
%%%
\par
The candidate circulation periods for other sub-sequences were also
successfully cross-validated by examining longitude-longitude correlation
maps\footnote{For the sub-sequence D$_1$, after examining the
longitude-longitude correlation maps, the low frequency feature is treated
as the first sub-harmonic of that corresponding to the candidate $P_4$.
Since the spectral power corresponding to the candidate $P_4$ is not
significant, the subsequent results from correlation analysis of this
sub-sequence should be seen with caution. However, $P_4$ directly
corresponds to the observed low frequency feature for the other three
sub-sequences, and our overall conclusions are in no way biased by our
analysis of the sub-sequence D$_1$.}.
Figure~\ref{polar_maps_all} shows the variety in polar emission
maps constructed using these candidate $P_4$ values for the respective
sub-sequences. Table~\ref{mod_summary} summarizes the relevant parameters
for these different sets, namely, the number of pulses in each of the
sub-sequences, the candidate $P_4$ along with the number of sub-beams
in the mapped carousels corresponding to the two cones.
Pending the considerations in the next section, in
Table~\ref{mod_summary} we have denoted the candidate $P_4$ as
$P_{\rm LF}$, i.e., the period deduced using the low fluctuation
frequency feature. Similarly, we have denoted $P_3$ as $P_{\rm HF}$
(period deduced using high fluctuation frequency feature).
\par
Visual inspection of the emission maps, even after appropriate smoothing,
could not help in reliably estimating the number of sub-beams, except
for the sub-sequence A$_1$. In
an attempt to determine the number of sub-beams in a more systematic way,
azimuthal sequences averaged within ranges of radii corresponding to the
outer and the inner carousels were computed by sampling the emission maps
at uniform intervals of magnetic azimuth. These sequences were then Fourier
transformed to estimate the periodic azimuthal spacing between the sub-beams,
and hence the number of sub-beams in the carousel. Note that the above
procedure is prone to under-estimate the number of sub-beams, when the
intensities of individual sub-beams are highly non-uniform. Further,
the sub-sequences of small lengths (e.g., B$_1$) are prone to under-sample
the emission map (especially towards larger magnetic colatitudes), and
hence might mislead to a wrong number of sub-beams. Due to these reasons,
our estimates of the number of sub-beams are generally less reliable
(particularly for sub-sequences other than A$_1$), but are
listed in Table~\ref{mod_summary} for the sake of completeness.
%%%
%%=======================================================================
%%%-------------------------------------------------------------------------
\begin{figure*}
\centering
\subfigure[]{
 \includegraphics[width=0.5\textwidth,angle=-90]{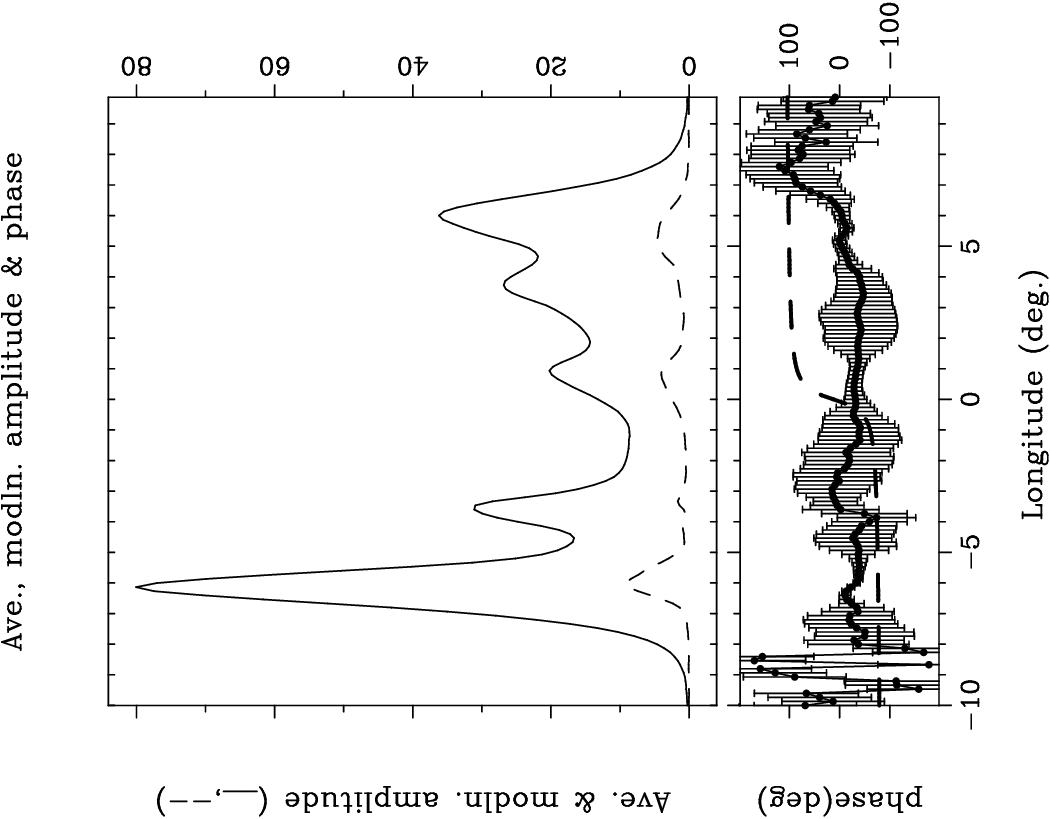}}
\hspace*{2mm}
\subfigure[]{
 \includegraphics[width=0.5\textwidth,angle=-90]{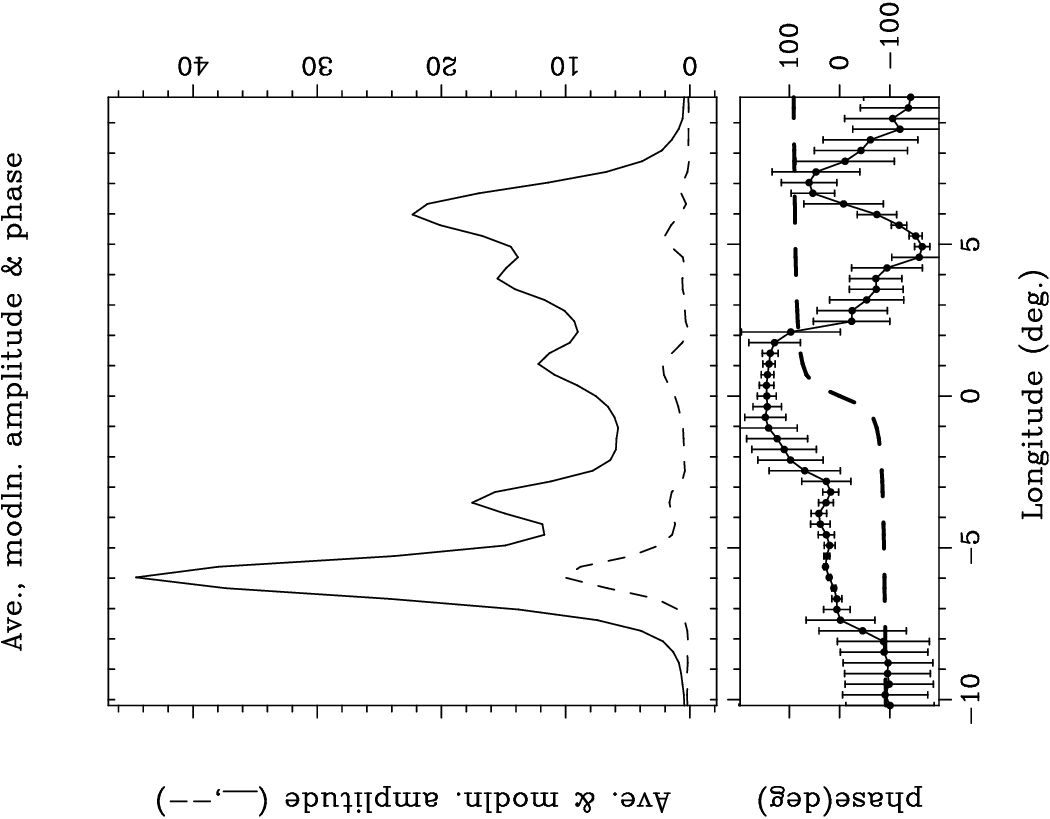}}
\vspace*{2mm}
\subfigure[]{
 \includegraphics[width=0.5\textwidth,angle=-90]{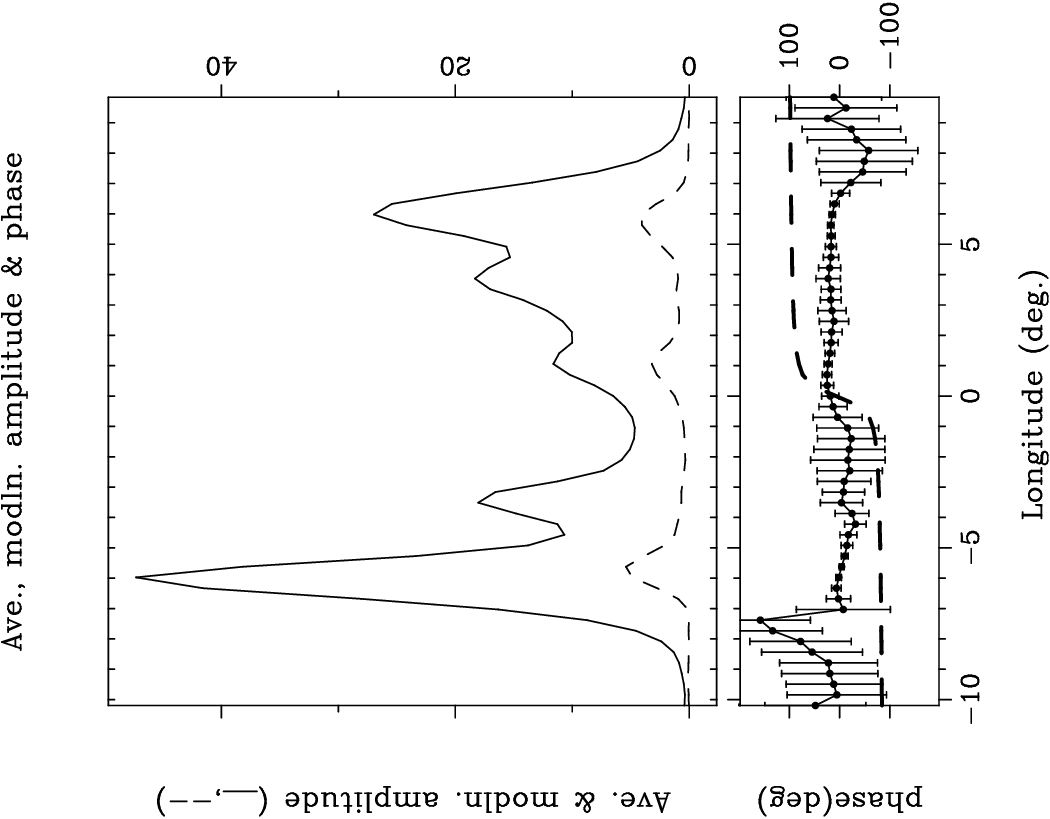}}
\hspace*{2mm}
\subfigure[]{
 \includegraphics[width=0.5\textwidth,angle=-90]{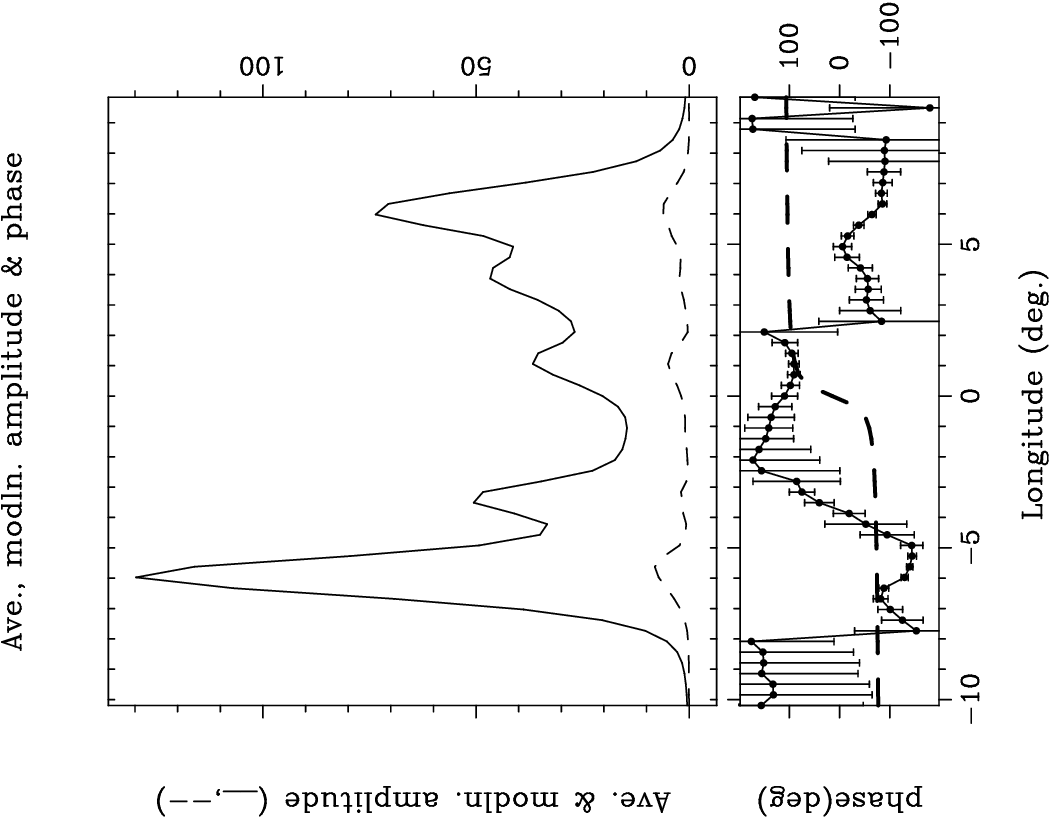}}
 \caption{\textsl{Tertiary modulation Amplitude and Phase:}
In each of the 4 sub-figures, the upper panel shows the stokes-I
average profile (solid line), and the amplitude profile of the modulation
presumed to be due to the carousel rotation (dashed line). The bottom panel
shows the observed modulation phase profile with $\pm1\sigma$ error-bars
($\pm2\sigma$ for the last one)
as a function of pulse-longitude, along
with the modulation phase sweep expected from the carousel model (dashed line).
The 4 sub-figures, (a), (b), (c) and (d), correspond to
the sub-sequences A$_1$, B$_1$, C$_1$ and D$_1$, respectively.}
 \label{modphase}
\end{figure*}
%%%-------------------------------------------------------------------------
%%=======================================================================
%%%-------------------------------------------------------------------------
%%%-------------------------------------------------------------------------
\section{The modulation-phase and the carousel model}\label{sect_modphase}
The sweep of the modulation phase across the pulse profile provides an
important consistency check to assess if the observed fluctuation spectral
feature is indeed a manifestation of the carousel rotation.
Figure~\ref{modphase} presents the observed modulation phase profiles
corresponding to the observed low frequency features, along with that
expected from the carousel model, for each of the four sub-sequences.
A glaring mismatch between the observed and expected profiles is obvious.
Note that a possible relative phase offset together with a
co-latitudinal overlap between the individual carousels, as well as
a different (i.e., non-carousel) origin of the core-component, might
cause the modulation phase profile to deviate from the expected trend.
However, such deviations should be least prominent at longitudes
where the received flux density is expected to originate mostly from
one of the cones, e.g., peaks of the leading and trailing components,
as well as outskirts of the profile. Apparent lack of consistency between
the observed and the expected modulation phase profiles, even in such
longitude regions, necessitates consideration of other possibilities.
\par
Nulling may also contribute to deviations from the expected modulation
phase profile. Pulsar B0809$+$74, due to its very stable
drifting and frequent short nulls, is the only source for which the
effect of nulls on the drift rate has been studied in detail. A
common picture which has slowly emerged from various studies of this
aspect \citep[e.g.,][]{Page73,Unwin78,LA83,van02} is that a perturbation
at the onset of a null turns off the emission and abruptly alters the
drift rate, which then recovers exponentially to its normal value.
However, the actual behavior of the drift during the nulls remains still
unclear. For example, either a complete cessation of the drifting after
a possible lag from the null onset, or a slowed-down drifting active
throughout the null duration, is possible. For B0809$+$74, the speeding-up
from slow/ceased drifting to normal starts during or at the end of the
null episode, and the recovery time is proportional to the null-length
\citep{LA83}. Possible deviations from this picture, when we consider
other pulsars, are not ruled out.
\par
To explore if the apparent lack of stability in the presumed carousel
patterns of B1237$+$25 is caused by the nulls, we have carried out a
systematic search, for what otherwise may be a stable emission pattern,
by modeling the effect of nulls in a number of different ways. We first
identified the null-pulses by examining the pulse-energy
histograms\footnote{We identified all the pulses having energy below
a chosen threshold as null-pulses. For this purpose, we examined the
pulse-energy histograms of the four pulse sequences, and selected
appropriate thresholds to include all the pulses under the narrow
distributions centered at 0 as null-pulses. Note that the distributions
of `normal' and null-pulses partly overlap. However, the number of pulses
which could be mistaken as null-pulses, or the null-pulses which could
be missed because of this overlap are estimated to be, on average,
less than or about $0.5\%$ of total number of pulses (and $\sim$1--2\%
of the null pulses).}.
Using the list of null-pulses, we modified the pulse number tag of each
pulse by a number of trial correction offsets (details of which are given
below), and explored whether a stable carousel pattern could be obtained.
We have assumed, in our following discussion, that the carousel of sparks
recovers to a common and stable rotation rate after every instance of
perturbation caused by nulls, and the total time spent in the perturbation
and recovery phase is dependent on the cumulative (from start of sequence)
null-duration.
\par
Each affected section of the pulse-sequence (i.e., the section
corresponding to irregular modulation) can be considered as starting
from a given null and extending till the time when both, (a) the stable
drift rate is attained, as well as, (b) the tertiary modulation phase
is same as that before the null. If such sections are removed, the remnant
sequence is expected to have coherent modulation due to carousel rotation.
To effect such a correction, the time of arrival of
n$^{\rm th}$-pulse ($T_{\rm n}$) is modified by the correction offset
$\Delta T_{\rm offset,n}$, which depends on the cumulative time spent
in all the irregular modulation phases till the n$^{\rm th}$-pulse,
in the following way:
$T_{\rm n,modified} = T_{\rm n} + \Delta T_{\rm offset,n}$.
We explored several models of this correction offset, incorporating
different possible linear dependences of the perturbation and recovery
durations on the null-extent.
We used the following trial forms of $\Delta T_{\rm offset,n}$, with
$\delta T$ as the parameter to be varied in fine steps:
(1) $N_{\rm null-durations,n}\times\delta T$,
(2) $N_{\rm nulls,n} \times \delta T$,
(3) $N_{\rm nulls,n}\times P + \delta T$, and,
(4) $N_{\rm nulls,n}\times(P + \delta T)$,
where $N_{\rm null-durations,n}$ is the total number of contiguous
null-durations and $N_{\rm nulls,n}$ is the total number of null-pulses,
encountered till the $n^{\rm th}$ pulse. In the first two models,
the perturbation to the carousel (or to the carousel rotation) and start
of the recovery phase are assumed to be instantaneous, while for the last
two models, perturbation time is less than the null-duration and the
recovery starts only when the null-duration is over.
The recovery time is assumed to be constant for the models (1) and (3),
while it is proportional to the null-length for the other two models.
\par\noindent
\textbf{\textsl{Investigations of the fluctuation spectrum:}}
Each of the above modeled correction offsets, with the variable parameter
$\delta T$, was applied to the time sequences corresponding to the first
and last components of the average profile.
The resultant time sequences (which are now, in general, non-uniformly
sampled) were Fourier transformed. The fluctuation spectra were examined
for compact features, which might emerge when coherence in the modulation
is improved via modelled correction for nulls, and which might correspond
to the carousel rotation time $P_4$. Such a feature may also be accompanied
by symmetric side-lobes due to uneven sampling. A compact secondary modulation
feature was also looked for, although it may not be always apparent. Further,
the feature corresponding to the carousel rotation should be present in
the fluctuation spectra of both the components. Hence, this criterion was
used to filter out the features not related to the carousel rotation.
Note that the profile mode changes might also contribute to worsening
the coherence in the modulation.
To explore the effects of null occurrences on the sub-pulse modulation in
isolation, this analysis was also applied to sub-sequences free from
profile mode changes (assessed via visual inspection), in addition to
the full-length pulse sequences. None of the above modeled correction
offsets helped in finding any carousel rotation feature common to both
the components.
%%%-------------------------------------------------------------------------
%%%-------------------------------------------------------------------------
\begin{figure*}
\centering
  \subfigure[The Polar Correlation map]{\includegraphics[width=0.44\textwidth,angle=-90]{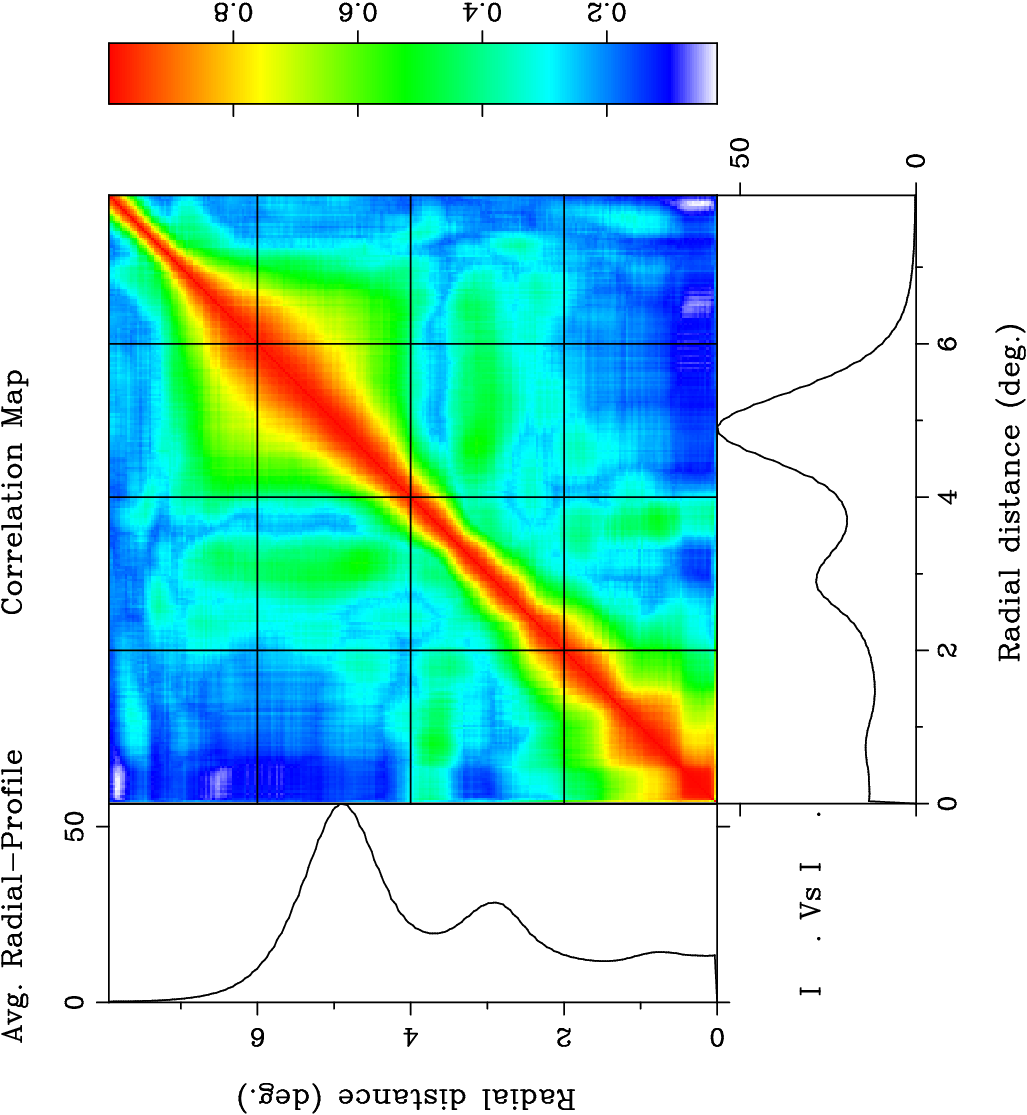}}
  \hspace*{2mm}
  \subfigure[The Polar Azimuthal-Shift map]{\includegraphics[width=0.44\textwidth,angle=-90]{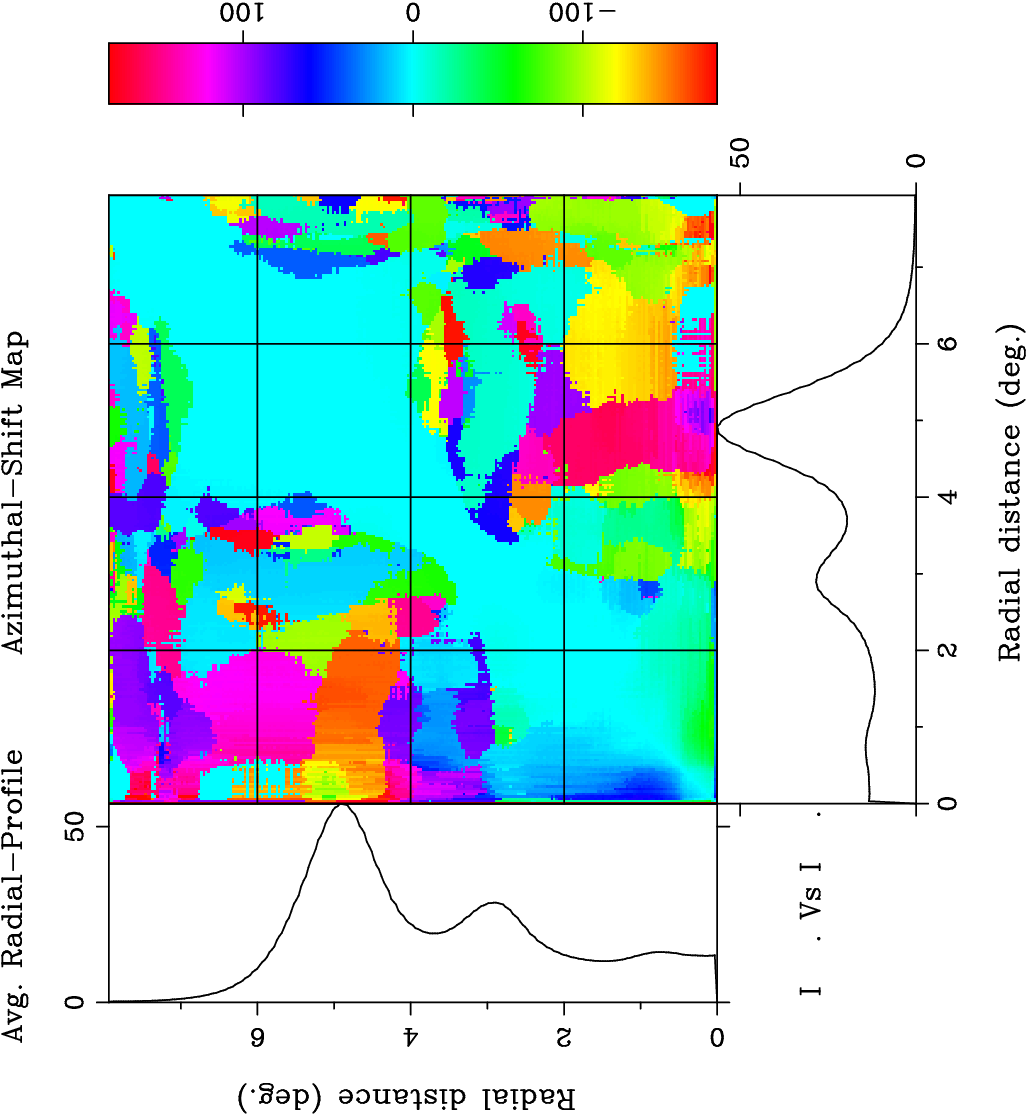}}
  \caption{\textsl{The Polar Correlation and Azimuthal-shift maps:}
\textbf{(a)}
The central panel shows the map of the maximum correlation coefficients
(normalized) computed between azimuthal intensity fluctuations at all
possible pairs of radial distances (i.e., magnetic colatitudes) in the
emission map shown in Figure~\ref{polar_maps_all_a}.
The bottom and left-hand side 
panels show the average radial profiles.
\textbf{(b)}
The central panel shows the map of magnetic-azimuthal shifts (in degrees)
corresponding to the maximum correlation coefficient values plotted in (a).
The bottom and left-hand panels show the average radial profiles.}
 \label{corr_maps}
\end{figure*}
%%%-------------------------------------------------------------------------
%%%-------------------------------------------------------------------------
\par
A different category of nulls --- pseudo-nulls --- has been identified
recently \citep{HR07,HR09}, wherein the nulls represent chance positioning
of our sightline across the minima (i.e., in between the sub-beams) in
the carousel(s), instead of the actual cessation of the emission. A few
examples of the pulsars which are now known to exhibit pseudo-nulls are:
B0834$+$06 \citep{RW07}, B1133$+$16 \citep{HR07} and B2303$+$30 \citep{RWR05}.
The pseudo-nulls in B1237$+$25 would have been periodic and easy to identify,
had the tertiary modulation been stable, which is unfortunately not the case.
In any case, pseudo-nulling is less likely in B1237$+$25 because the two
presumed carousel systems are azimuthally offset (as we show in the next
section) in such a way that the probability of positioning our sightline
simultaneously passing through minima in both the carousels would be
even smaller than that for random chance occurrence\footnote{If P(A)
and P(B) are the probabilities of `empty' sightline traverses through
the two carousel sub-beam systems individually, then the probability
of an `empty' sightline traverse simultaneously through the two carousels
would be P(A)$\times$P(B).}. Probability of such null occurrence would
further reduce, if indeed the core-component were to have a different
origin than the conal components.
Nevertheless, given the secondary modulation period of $\sim2.7$ P and
finite physical widths of sub-beams, apparent duration of any possible
pseudo-null is unlikely to exceed one pulse period.
So, considering the extreme case, we assumed all the one period long
nulls as pseudo-nulls (and treated them like non-null pulses).
With the remaining longer duration nulls, our re-explorations,
using the modeled correction offsets and subsequent investigations
described above (previous two paragraphs), also did not result in any
carousel rotation feature consistent in fluctuation spectra of both
the components.
\par
It is possible that the perturbation time and/or the recovery time varies
from one
instance of nulling to another, and does not have any particular dependence
on the time the pulsar spends in null-phase. Even in such a case, if the
nulls are randomly distributed across the observed sequence, the observed
$P_4$ may deviate from the actual value depending on the fractional time
spent in nulls. As seen from Table~\ref{mod_summary}, $P_4$ might appear
to have an inverse dependence on the null-fraction, however, given our
limited statistics it will be too premature to infer any details based
on this apparent dependence. In the absence of any useful guidance
forthcoming from the above mentioned modeling of effects of null-occurrences
in identifying a stable carousel rotation period, and hence in disentangling
the corresponding effects in the modulation phase profile, we proceed with
the presumed association between the observed low-frequency features and
the carousel rotation, for various sub-sequences.
%%%
%%%=========================================================================
%%%=========================================================================
\section{Correlation properties of the emission patterns}\label{sect_cones}
The emission patterns presented in section~\ref{section_maps} clearly show
presence of two carousels corresponding to the two pairs of conal components
and a diffuse pattern corresponding to the core region.
To explore any inter-relationship between these patterns,
the radial profiles from the
emission maps were obtained at uniform intervals (1$^\circ$) of magnetic
azimuth, and then this whole azimuth sequence of radial profiles
was subjected to correlation analysis over the full span of azimuthal
shifts. The maximum correlation for each pair of radii (i.e., magnetic
colatitude) and the corresponding azimuthal shift was found out, and
plotted in two maps: ``The Polar Correlation map'' and ``The Polar
Azimuthal-Shift map'', respectively. Figure~\ref{corr_maps} shows
these maps computed for the azimuthal sequences from the emission map
presented in Figure~\ref{polar_map}
(i.e., for the sub-sequence A$_1$)\footnote{The diagonal line in the
correlation map, from top right to the bottom left, corresponds to the
``zero''-azimuthal shift autocorrelation, and by definition, a normalized
correlation coefficient of unity.}. In these maps, the $1\sigma$
uncertainty in the (normalized) correlation coefficients is estimated
to be about 0.1. A number of points to be noted from these correlation
maps are as follows.\\
1.) Significant correlation (maximum of nearly 60\%) is seen between the
two conal rings at a non-zero azimuthal shift. Also, there is significant
correlation (up to about 50\%) between the diffuse pattern corresponding
to the ``core''-emission and the region \emph{between} the two
conal rings.\\
2.) The ``boxy'' patterns symmetric about the autocorrelation track
corresponding to the peaks of the outer cone and the core-emission
are due to the finite radial extent of the corresponding patterns
(for the outer cone, it is centered slightly towards the outer edge
of the cone). It is quite surprising that there is no apparent ``boxy''
pattern corresponding to the inner cone.\\
3.) In the correlation ``box'' corresponding to the core-emission, the
azimuthal-shift corresponding to the maximum correlation increases
smoothly if we go perpendicular to the diagonal line corresponding to
the auto-correlation. This trend continues along the outward radial
direction till we reach the inner cone's peak. For the outer-conal
region, this trend is not so prominent and visible only towards the outer
edges. These systematic variations in azimuthal-shift might be indicating
a small azimuthal twist of the pattern as we go away from the magnetic
axis, prominently seen only at low magnetic co-latitudes.
%%%=========================================================================
\begin{deluxetable}{cccccc}
\tabletypesize{\footnotesize}
\tablecolumns{6}
\tablewidth{0pt}
\tablecaption{Correlation between the \emph{outer} and the \emph{inner} conal emission patterns: Summary of parameters.}
\tablehead{
   \colhead{Sub-seq.}                             	&
   \multicolumn{2}{c}{Correlation Function}			&
   \colhead{}							&
   \multicolumn{2}{c}{Phase-gradient fitting} \\
   \cline{2-3} \cline{5-6} \noalign{\smallskip}
   \colhead{}                                    	&
   \colhead{$\rho_{\rm max}$(\%) }                     	&
   \colhead{$\Delta\phi_{\rm az}$(\mdeg)}		&
   \colhead{}						&
   \colhead{$\Delta\phi_{\rm az}$(\mdeg)}		&
   \colhead{$\Delta\tau$(P)}
	}

\startdata
%---------------------------------------------------------------------------
A$_1$	&$45\pm09$ &$+10\pm2.5$ & &  $-30.3\pm0.3$ & $-2.39\pm0.03$ \\
%---------------------------------------------------------------------------
B$_1$  	&$55\pm12$ &$+13\pm3.5$ & &  $+13.1\pm0.4$ & $+0.66\pm0.02$ \\
%---------------------------------------------------------------------------
C$_1$  	&$41\pm09$ &$+09\pm3.5$ & &  $+10.0\pm0.3$ & $+0.94\pm0.03$ \\
%---------------------------------------------------------------------------
D$_1$  	&$35\pm09$ &$+10\pm4.5$ & &  $+09.6\pm0.4$ & $+0.62\pm0.03$ \\
%---------------------------------------------------------------------------
\enddata
\tablecomments{The maximum percentage correlation is denoted by
$\rho_{\rm max}$, and the corresponding magnetic azimuthal shift,
$\Delta\phi_{\rm az}$ (in degrees), is the amount by which the
pattern corresponding to the outer cone lags behind that corresponding
to the inner cone. The uncertainty in $\Delta\phi_{\rm az}$ corresponds
to 68\% confidence interval around the $\chi^2$-minimum. $\Delta\tau$
is the absolute time-delay equivalent to the above shift in units of
pulsar rotation periods.}
\label{corr_summary}
\end{deluxetable}
%%\clearpage
%%%=========================================================================
%%
\par
Similar correlation patterns, specifically showing the correlation between
the two conal patterns, as well as between the core and inter-conal region,
were observed in the correlation and azimuthal-shift maps constructed for the
other sub-sequences discussed above.
A noticeable difference found between various correlation maps is that
the locations of enhanced correlation with the core-region are seen to
vary across the radial direction, in some cases even extending up to the
regions corresponding to the two conal patterns.
\par
To present a quantified measure of the correlation parameters, azimuthal
sequences averaged over narrow radial sections, each about half a degree
to one degree wide in colatitude, centered at the plateau corresponding
to significant correlation between the two conal rings, were correlated.
The resultant correlation function provided us the maximum correlation
and the corresponding azimuthal shift between this pair of patterns.
To measure the shifts more
precisely, the cross-spectrum was examined to select spectral features
with amplitudes above a chosen threshold, and the corresponding phase
profile was fitted with a
linear gradient. The azimuthal shifts corresponding to the best-fit
phase-gradients (determined by minimizing $\chi^2$) and those obtained
from the correlation function, along with the observed maximum correlation
between the above pair of averaged azimuthal sequences, are presented
in Table~\ref{corr_summary}. Significant correlation between the
two patterns is clearly evident consistently across various sub-sequences.
Further, the azimuthal shifts estimated using the above two methods are
consistent with each other, except for the sub-sequence A$_1$. However,
we should note that the phase gradients are fit over \emph{limited}
parts of the cross-spectrum, and hence the corresponding estimates of the
azimuthal shifts, are subject to possible aliasing due to the secondary
modulation. Note that the sub-beam spacing for this sub-sequence is
about $20\mdeg$, and hence, a measured shift of $-30\mdeg$ is consistent
with an actual shift of $+10\mdeg$ as indicated by the correlation function.
With this understanding, we note that the azimuthal shifts estimated for all
the sub-sequences are consistent with each other, i.e., all of them are
around $+10\mdeg$.
\par
We have also seen the emission pattern of the core-region to be correlated
with that of the ``inter-conal'' region. However, the significance and the
radial location of the enhanced correlation is not consistent across the
sub-sequences.
Hence, we have limited our further discussion only to the correlation
between the inner and outer conal emission patterns.
%
%%
%%%=========================================================================
%%%=======================================================================
\section{Discussion}\label{sect_discussion}
%%%
We have presented a number of maps of the underlying emission sub-beam
configurations of the pulsar B1237$+$25, each reconstructed using the
best-fit candidate $P_4$ value inferred by analyzing the rich sub-pulse
modulations in the corresponding single pulse sequences from this
M-category star.
Like many other pulsars, the inferred $P_4$ values are several
times larger than that predicted by the R\&S model for this pulsar
($\approx 3.4~P$). \citet{GS00} have recently modified the R\&S model.
Their ``modified carousel model'' predicts the circulation time for
B1237$+$25 to be in the range 24--32~$P$ (corresponding to the sparks
associated with the outer cone generated at a distance 0.7--0.8 times
the polar cap radius away from the pole). This predicted range is
in complete agreement with the $P_4$ inferred for the sub-sequence A$_1$.
\par
The sub-beam configurations vary across different
sub-sequences, and no firm (harmonic) relationship is apparent between
the various inferred $P_4$ values, suggesting that the underlying
sub-beam configurations are not stable over long durations.
The short scale variations in the emission maps, along with a very
low-Q feature corresponding to the secondary modulation, indicate that
the configurations are not quite stable even within the interval of short
sub-sequences considered here.
Further, neither the current understanding of the physical
mechanism responsible for subpulse modulation nor any observational
evidence from other pulsars known to exhibit subpulse modulation suggests
a large spread like that apparent in our inferred circulation times.
However, such a large spread may be understood if the underlying
carousel of sub-beams of this pulsar is perturbed on time-scales
shorter or comparable to a few times the circulation periods. In such
a scenario, the apparent low frequency feature in specific sub-sequences
would represent a chance peak resulting from an unavoidable beat
between the carousel circulation and the perturbing modulation. And
if our selection of sub-sequences is biased to those showing a reasonably
isolated low-frequency feature, then we may be preferentially looking
at such manifestations of the beats. The perturbing modulation of
relevance here would be necessarily in the form of a phase modulation,
since it would shift the apparent (perturbed) modulation frequency
systematically away from its true value. Any perturbing amplitude
modulation, on the other hand, would result in symmetric sidebands,
without any shift locally in the centroid of the perturbed modulation
features. If many random realizations of such perturbations were to
be available, even if in form of phase modulations, then an ensemble
average of the possible shifts could be expected to be close to zero.
Therefore, the actual circulation time could be estimated from the
average location of the beat frequency feature, obtained from a
statistically significant number of sub-sequences providing a
distribution of such chance beat frequencies. Such an estimate from
our set consisting of only 4 sub-sequences suggests a circulation
period of $25.95\pm0.10$~$P$.
Further note that the effects of the above perturbing modulation
would also reflect in the frequency feature corresponding to $P_3$.
The observed range of $P_3$ along with the observed number of
sub-beams for the sub-sequence A$_1$ suggests a dispersion of about
$13~P$ in $P_4$, which is well consistent with the observed variation
in $P_4$ (the difference between the maximum and minimum $P_4$
estimates is about $15~P$; see Table~\ref{mod_summary}).
Although the above scenario involving perturbing modulations implies
that our inferred $P_4$ values, from specific sub-sequences, would
deviate from the actual circulation time, the correlation between
the inner and outer patterns and the inferences drawn therefrom
would still hold their significance.
\par
Nulling can cause perturbation in the carousel rotation, and hence in
the sub-beam drift rate, and if these are not accounted for, the maps
in the rotating frame would show irregular patterns, such as those
observed. The large range of the sub-beam circulation periods, inferred
from various sub-sequences, also prompted us to explore if it has any
relationship with the fraction of time the pulsar remains in null-mode.
There is indeed a hint of an inverse relationship between $P_4$ and
null-fraction. However, the apparent dependence is not statistically
significant enough to draw any useful inference. We tried to model the
effects of nulls as perturbation in the carousel rotation followed by
recovery to a stable rotation rate, and searched for a stable carousel
of sub-beams. The negative results from our modeling of the time spent
in the perturbed and recovery phases in 4 different ways suggest that
the recovery time does not depend linearly on the preceding null-duration,
and is possibly random or has a non-linear dependence. Even for suitably
long null-free sub-sequences, we did not see an indication of a stable
emission pattern through our spectral and correlation analysis. More
importantly, the observed profile of modulation phase across the
pulse-window deviates significantly from, and appears to have little
correspondence with, that predicted by the carousel model. The deviations
appear to be random across various sub-sequences, and might be understood
as due to lack of stability in the emission patterns, unless the mismatch
is actually pointing to failure of the standard carousel model.
%%%%
\par
The major question we have been trying to address is: do the ``inner''
and ``outer'' cones share the same seed bunch of particles for their
excitation ? This question had been raised first by \citet{Rankin93a},
followed by `indirect' evidences by \citet{GG01} which, in case of
pulsar B0329$+$54, seem to support the view that the emission in
multiple cones is associated with the same set of magnetic field
lines but at different altitudes. The polar emission maps of the
pulsar B1237$+$25 have provided us a comprehensive way to study a possible
relationship between the sub-beam patterns responsible for the emission
in the inner and outer cones. As evident from the polar correlation map
for the sub-sequence A$_1$ discussed in previous section, the patterns
associated with the two cones are significantly correlated with each
other, and consistently so, for the other sub-sequences discussed above
(see Table~\ref{corr_summary}). This correlation between the two patterns
provides, in our view, a `direct' evidence for the same seed pattern of
``sparks'' being responsible for emission in the two cones. It is worth
noting here that the actual correlation between the two patterns would
probably be even higher, if (a) the observed lack of stability of the
sub-beam configurations, and (b) the possible overlap between the radial
extents of the carousels, combined with the fact that there is a relative
azimuthal shift between them, may have distorted the mapped patterns.
\par
The sub-beam patterns in the two conal rings are found
to be offset (relative to each other) by a magnetic azimuthal shift of
about $10\mdeg$, consistently for all the sub-sequences.
An early indication of such an offset comes from \citet{HW80},
wherein they show relative phase offsets between the subpulse modulations
under the conal components of B1237$+$25. The sense of subpulse ordering
they had noticed between the inner and outer conal components (see their
figure 2), with delays of about 1 period, is consistent with the
corresponding emission patterns having a stable phase offset, as we found
above.
Note that an azimuthal offset between the two orthogonal
polarization mode (hereafter OPM) emission elements of the conal
sub-beams \citep[as suggested by][]{RR03}, combined with significantly
different fractions of two OPM powers in the two cones \citep{SRM13},
might also manifest an azimuthal shift between the two patterns. However,
consistency of the observed shift would require similar proportions of
the two OPMs across various sub-sequences.
In any case, the analysis presented in Appendix shows that
the emission patterns corresponding to the two OPMs do not have
any relative azimuthal offset between them.
Hence, the observed magnetic azimuthal shift does not have its
origin in complexity of the OPMs.
\par
For our sub-sequence A$_1$, the above azimuthal offset of about $10\mdeg$
implies that the azimuthal positions of sub-beams in the inner carousel
fall in between those of the outer ring sub-beams\footnote{The 18 sub-beams
in either of the two rings (for sub-sequence A$_1$, where number of sub-beams
could be determined with some certainty) implies a sub-beam azimuthal spacing
of $20\mdeg$ ($360\mdeg/18$).}.
\citet{Bh09} interpreted a similar finding in pulsar B0818$-$41 to
be consistent with a maximal packing of sparks on the polar cap, as
proposed in the ``modified carousel model'' \citep{GS00}. However,
we note that this model considers a characteristic spark dimension
that is equal to the typical distance between the sparks\footnote{The
characteristic spark dimension, as well as the typical distance between
the sparks, is considered to be equal to the polar cap height.}.
Such maximally packed polar cap would naturally lead to a \textit{smaller}
number of sub-beams in the inner ring, which is in direct contradiction to
the same number found in the two rings of B1237$+$25 as well as B0818$-$41.
Hence, the two apparent carousels in these pulsars, and perhaps in all the
double-cone pulsars with regular drift, can not be explained by the
modified carousel model.
For our observations to be consistent with two independent systems of
sparks (and corresponding sub-beams) the spark's characteristic dimension,
and/or the inter-spacing between the sparks, in the inner ring need to
be decreased appropriately, with a physical justification to do so.
To the best of our knowledge, the existing carousel models do not
suggest any such dependence of spark's size/inter-spacing on the
magnetic colatitude. In the absence of a justification for the same
number of sub-beams in the two carousels, the observed correlation
between the intensity variations of sub-beams in the two patterns
of B1237$+$25 also remains unexplained in such models.
\par
Further, the emission in the outer and the inner cones is believed to
originate from distinct altitudes with the outer cone
exhibiting the RFM and the inner cone originating from a nearly
fixed height \citep[see, e.g.,][]{Rankin93a,Rankin93b,MR02}.
Using the aberration-retardation considerations, the 327 MHz emission
altitudes of the outer and the inner cone of B1237$+$25 are estimated
to be $340\pm79$ and $278\pm76$ km, respectively. However, by including
the rotational sweep-back effects of the field lines which become
important at altitudes $\lesssim 1\%$ of the light cylinder radius
\citep{DH04}, we estimate the emission altitudes of the outer and the
inner cone to be in the ranges 690--740 and 610--670 km, respectively.
Noting these estimates, along with the observed
correlation between the emission patterns in the two cones, it seems
an obvious and plausible interpretation that the same underlying pattern
is responsible for the same frequency emission at two distinct altitudes.
In this ``multi-altitude emission'' picture, the inner cone emission comes
from a lower altitude than the emission in the outer cone, along the
same bunches of magnetic field lines.
However, to explain the lack of evolution of the inner conal component
separation with frequency, additional considerations regarding the
emission or propagation of waves \citep[e.g., refraction of rays in the
open field line region,][]{BA86} would be required.
\par
The light travel time between the estimated emission altitudes of the two
cones ($<1$ ms) is negligible when compared to the delays equivalent to
the azimuthal offsets (see last column of Table~\ref{corr_summary}),
and can not account for the observed offset of about $10\mdeg$.
Hence, the large offset between the two patterns, in the above
``multi-altitude emission'' picture,
implies a twist in the emission columns, possibly associated with a twist
in the magnetic field lines, between the two different emission altitudes.
A similar scenario was invoked by \citet{RSD03} to explain the
mode-switching phenomenon in B0943$+$10, wherein the emission columns
(coupled with the magnetic field) are twisted progressively at larger
distances in the relatively weaker magnetic field regions.
\par
B1237$+$25 is the first pulsar wherein the inner and outer cone components
were noticed to be modulated with a constant phase offset \citep{HW80}.
A few more double-cone pulsars have been found to exhibit such
phase-locked modulations \citep[B0826$-$34, B0818$-$41,
B1039$-$19, B1918$+$19; see][respectively]{Gupta04,Bh07,Bh11,RWB13},
and no counter-examples are known so far --- suggesting the phase-locking
to be a common feature in double-cone pulsars with regular drift
\citep[as already noted by] []{RWB13}.
Findings in B1237$+$25 and B0818$-$41 then suggest the origin
of the phase-locked modulation to be the same number of sub-beams in the
two \emph{apparent} carousels circulating around the magnetic axis with
the same period. We again note that such configurations can not be explained
by the modified carousel model.
Multi-altitude emission, as suggested by the correlated sub-beam intensity
variations in the two conal patterns of B1237$+$25, should also explain
the origin of the inner cone in other double-cone pulsars.
\par
The other important question we have tried to ask and address is about the
origin of the core-component. Although the core-component does not show any
signature of corresponding secondary modulation feature seen in the
conal-components ($\sim$ 0.37 c/P), the tertiary modulation feature
is present throughout the profile and prominently so in the ``core-region''
(i.e., the region covering the core-component and the region prior to it;
see Figure~\ref{fluc_summary_a}). This observation is true
for all the sub-sequences explored above.
Although the origin of the core component is believed to be
different from that of conal components,
the above evidence of core-region sharing the same tertiary modulation
feature as the two conal sub-beam patterns, provides a different point of
view: at least 
some part of the core emission might also originate from a compact and
`further in' sub-beam pattern. The diffuse nature of the pattern, i.e.,
the absence of any discrete sub-beam structure, would then explain the
apparent lack of secondary modulation feature in the fluctuation spectrum
as it would be attenuated by smoothing due to the finite sub-beam width.
We note that one of the two OPM portions of the core region
exhibits intensity dependent aberration/retardation effects
\citep{SRM13} that are interpreted as a cascade or amplifying process
along the magnetic axis. These processes may also contribute to
smoothing, and might as well to deformation, of an otherwise discrete
and compact sub-beam pattern.
The ``white'' fluctuation spectrum for the core-emission seen in many
cases \citep[specifically for core single profile stars; ][]{Rankin86}
may be explained naturally if and as the diffuse nature of the pattern
approaches uniformity, and only the overall intensity fluctuations in
time survive. Given the above picture, it is tempting to suggest that
the conal as well as some part of the core
emission stem from the same seed pattern at
different heights in the magnetosphere, which in its unresolved form
contribute to
the core component and at a suitably higher distance gets
resolved out (longitudinally as well as latitudinally) in to a system
of discrete sub-beams that we see in the conal-emission. It should be
emphasized here that this picture of the core component would require
further considerations similar to what we need to explain absence of
RFM in the ``inner'' cone.
\par
The emission in the core-region appears to be correlated with that in the
inter-conal region (Figure~\ref{corr_maps}). However, the inconsistency
in location and significance of this correlation across various
sub-sequences limits our confidence in inferring any further detail
based on this correlation. None of our earlier inferences are however
affected by this limitation.
%%%
%%%
%%%
\section{Conclusions}\label{sect_conclusions}
With the specific aim to find out whether the outer
and inner cones of the pulsar B1237$+$25 share a common origin or not,
we have mapped and studied the underlying emission patterns for a number
of pulse sequences from this star.
The emission patterns corresponding to the outer and the inner cones
are found to be significantly correlated with each other, implying that
the emission in the two cones share a common seed pattern of sparks.
This main result is consistent with the same radio frequency emission
in the two cones, originating from a common seed pattern of sparks at
two different altitudes. More interestingly, the observed azimuthal
offset of about
$10\mdeg$ between the two conal emission patterns suggests a twist in
the emission columns, and most likely in the magnetic field geometry,
across the two different emission altitudes. We also addressed the
possibility that some part of
the core component --- contrary to common belief ---
also shares its origin with the conal counterparts, and a possible
presence of a compact, diffuse and \emph{further-in} carousel of sub-beams
can consistently explain the generally observed slow modulation, or
lack thereof, in the core component.
\par
The underlying carousel of sparks for this pulsar appears to seriously
lack stability over long durations. Even for the shorter sub-sequences,
the sweep of the modulation phase across the pulse longitudes
deviates significantly from that
predicted by the carousel model. Unless these deviations have
significant contributions from pulse-nulling, mode-changing and/or
inherent irregularities in the carousel on timescales much lesser
than the lengths of our individual sub-sequences, they pose a
serious challenge to the widely accepted standard carousel model.
%%%
%%%
\section*{Acknowledgments}
We are grateful to Joanna Rankin for making various of the above
discussed pulse-sequences (including the mode-separated versions
of one of the sequences) available to us.
We also thank Joanna Rankin for a critical review of our paper,
and for her constructive comments and suggestions which have helped
in improving the manuscript.
The Arecibo Observatory is operated by SRI International under a
cooperative agreement with the National Science Foundation,
and in alliance with Ana G. M\'endez-Universidad Metropolitana, and
the Universities Space Research Association.\\
{\it Facilities:} \facility{Arecibo Telescope}
%%%
%%%=========================================================================
%%%=========================================================================
%%%-------------------------------------------------------------------------
\begin{figure*}
\centering
\includegraphics[width=0.5\textwidth,angle=-90]{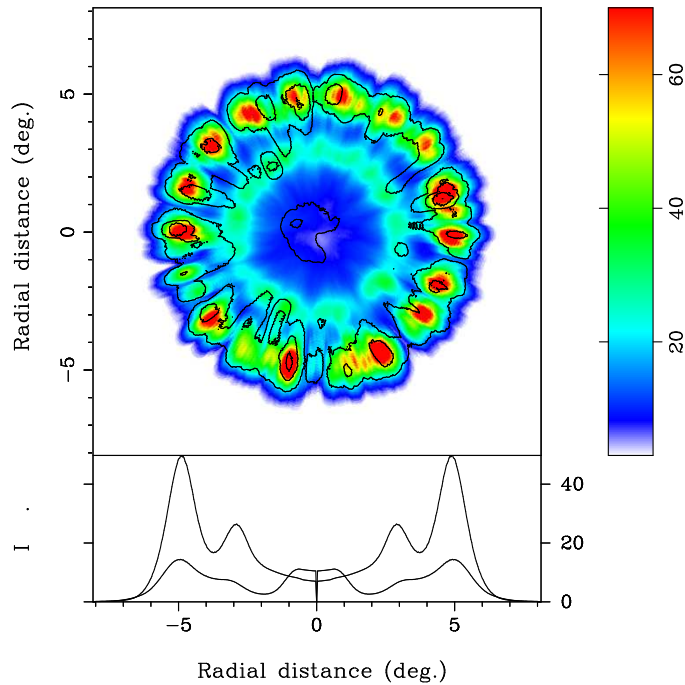}
\caption
{\textsl{Polar maps of the emission corresponding to the two OPMs:}
The color-image shows the polar emission map constructed
using only the PPM power from the sub-sequence A$_1$, and using the geometrical
parameters and circulation period value mentioned in the text. The overlaid
contours show the emission map constructed using only the SPM power from
this sub-sequence. We see that, on average, SPM emission follows the PPM
beams very well, without any offset in magnetic azimuth.
The bottom panel shows the average-intensity profiles for the two modes
(PPM: higher intensity profile; SPM: lower intensity profile),
as functions of the angular distance from the magnetic axis. Note that
the conal SPM power (contour-map) is much weaker compared to its PPM
counterpart --- it is only about 25--30\% (on average) of PPM power,
as indicated by profiles in the bottom panel.
}
\label{polar_maps_2p2s}
\end{figure*}
%%%-------------------------------------------------------------------------
\begin{appendix}
\section*{Polarization modal emission maps and their correlation properties}
Based on a study of average polarization (or rather depolarization)
properties of a few conal single and double profile pulsars, \citet{RR03}
proposed that the primary polarized mode (PPM) and the secondary polarized
mode (SPM) emission elements of the circulating sub-beams are offset from
each other in magnetic azimuth as well as colatitude.
The already observed magnetic azimuth offset between the
PPM and SPM beamlets of B0943$+$10 \citep{DR01} corroborated well
with this proposition, although no offset in magnetic colatitude was
found for this pulsar. \citet{SRM13} and \citet{RR03} have argued
that the OPM sub-beams are offset in
magnetic azimuth for B1237$+$25 (although their argument is based
on the phase analysis of the secondary modulation unlike the complete
carousel mapping analysis for B0943$+$10). Hence, it is important to
assess whether the observed azimuthal offset between the inner and
outer conal emission patterns has its possible origin in our usage
of mixed-mode pulse sequences.
\par
In Figure~\ref{polar_maps_2p2s}, we have plotted the polar emission
map reconstructed using only the PPM power from the sub-sequence A$_1$
as a color image, and that using only the SPM power as
contours\footnote{Other properties of the 2-way OPM-separated sequences
that are used for constructing these maps, can be found in \citet{SRM13}.}.
The geometrical parameters and the circulation period assumed are
same as mentioned in the main text for this sub-sequence. Note that
the inner conal components exhibit hardly any modulation in the
(already very weak) SPM emission, leading to structures only in
the outer ring in the corresponding contour map. The SPM emission
in the outer ring, on average, follows the PPM power sub-beams
very closely. No apparent azimuth offset between the OPM sub-beams,
and the PPM and total power resulting in virtually the same emission
maps (compare the image-maps in Figures~\ref{polar_maps_2p2s} and
\ref{polar_map}) already confirm that mixing of the OPMs have not
induced the observed offset between the two conal patterns. A formal
correlation analysis, as described in Section~4, using the PPM emission
map provides the same results as those with the total power map.
These results further confirm our conclusion above that the observed
azimuthal offset between the two conal emission patterns corresponding
to sub-sequence A$_1$ does not have its origin in complexity
of the OPMs.
However, we note that \citet{RR03} had found, using a very
old observation, a magnetic azimuthal offset between the PPM and SPM
sub-beams of B1237$+$25, and hence, the OPM sub-beams of this pulsar
may not always be azimuthally co-located.
\end{appendix}
%%%
%%%
%%%=========================================================================
%%%=========================================================================
%%%
%%%

%
%%%=========================================================================
%%=======================================================================
%%%-------------------------------------------------------------------------
%%%

\begin{thebibliography}{}
%
\bibitem[Backer(1970a)]{Backer70a}
Backer, D.~C.\ 1970a, Nature, 228, 42
%
\bibitem[Backer(1970b)]{Backer70b}
Backer, D.~C.\ 1970b, Nature, 228, 752
%
\bibitem[Backer(1970c)]{Backer70c}
Backer, D.~C.\ 1970c, Nature, 228, 1297
%
\bibitem[Backer(1973)]{Backer73}
Backer, D.~C.\ 1973, ApJ, 182, 245
%
\bibitem[\protect\citename{{Barnard} \& {Arons}, }1986]{BA86}
{Barnard}, J.~J., \& {Arons}, J. 1986, ApJ, 302, 138
%
\bibitem[Bhattacharyya et al.(2007)]{Bh07}
Bhattacharyya, B., Gupta, Y., Gil, J., \& Sendyk, M.\ 2007, MNRAS, 377, L10
%
\bibitem[Bhattacharyya et al.(2009)]{Bh09}
Bhattacharyya, B., Gupta, Y., \& Gil, J.\ 2009, MNRAS, 398, 1435
%
\bibitem[Bhattacharyya et al.(2011)]{Bh11}
Bhattacharyya, B., Wright, G., Gupta, Y., \& Weltevrede, P.\ 2011, American Institute of Physics Conference Series, 1357, 138 
%
\bibitem[\protect\citename{{Cordes}, }1978]{Cordes78}
{Cordes}, J.~M. 1978, ApJ, 222, 1006
%
\bibitem[\protect\citename{{Deshpande}, }2000]{Desh00}
{Deshpande}, A.~A. 2000, in ASP Conf. Ser. vol. 202, IAU Colloq. 177: Pulsar Astronomy -- 2000 and Beyond, ed. {Kramer}, M., {Wex}, N., \& {Wielebinski}, R., 149
%
\bibitem[\protect\citename{{Deshpande} \& {Rankin}, }1999]{DR99}
{Deshpande}, A.~A., \& {Rankin}, J.~M. 1999, ApJ, 524, 1008
%
\bibitem[\protect\citename{{Deshpande} \& {Rankin}, }2001]{DR01}
{Deshpande}, A.~A., \& {Rankin}, J.~M. 2001, MNRAS, 322, 438
%
\bibitem[\protect\citename{{Drake} \& {Craft}, }1968]{DC68}
{Drake}, F.~D., \& {Craft}, H.~D. 1968, Nature, 220, 231
%
\bibitem[\protect\citename{{Dyks} \& {Harding}, }2004]{DH04}
{Dyks}, J., \& {Harding}, A.~K. 2004, ApJ, 614, 869
%
\bibitem[\protect\citename{{Gangadhara} \& {Gupta}, }2001]{GG01}
{Gangadhara}, R.~T., \& {Gupta}, Y. 2001, ApJ, 555, 31
%
\bibitem[Gil \& Sendyk(2000)]{GS00}
Gil, J.~A., \& Sendyk, M.\ 2000, ApJ, 541, 351
%
\bibitem[Gupta et al.(2004)]{Gupta04}
Gupta, Y., Gil, J., Kijak, J., \& Sendyk, M.\ 2004, A\&A, 426, 229 
%
\bibitem[Hankins \& Wright(1980)]{HW80}
Hankins, T.~H., \& Wright, G.~A.~E.\ 1980, Nature, 288, 681
%
\bibitem[Herfindal \& Rankin(2007)]{HR07}
Herfindal, J.~L., \& Rankin, J.~M.\ 2007, MNRAS, 380, 430 
%
\bibitem[Herfindal \& Rankin(2009)]{HR09}
Herfindal, J.~L., \& Rankin, J.~M.\ 2009, MNRAS, 393, 1391 
%
\bibitem[\protect\citename{{Lyne} \& {Ashworth}, }1983]{LA83}
{Lyne}, A.~G., \& {Ashworth}, M. 1983, MNRAS, 204, 519
%
\bibitem[\protect\citename{{Maan} \& {Deshpande}, }2008]{MD08}
{Maan}, Y., \& {Deshpande}, A.~A. 2008, in AIP conf. ser. vol. 983, 40 Years of Pulsars: Millisecond Pulsars, Magnetars and More, ed. {Bassa}, C., {Wang}, Z., {Cumming}, A., \& {Kaspi}, V.~M., 103
%
\bibitem[\protect\citename{{Mitra} \& {Rankin}, }2002]{MR02}
{Mitra}, D., \& {Rankin}, J.~M. 2002, ApJ, 577, 322
%
\bibitem[\protect\citename{{Page}, }1973]{Page73}
{Page}, C.~G. 1973, MNRAS, 163, 29
%
\bibitem[\protect\citename{{Popov} \& {Sieber}, }1990]{PS90}
{Popov}, M.~V., \& {Sieber}, W. 1990, Sov. Astron., 34, 382
%
\bibitem[\protect\citename{{Radhakrishnan} \& {Cooke}, }1969]{RC69}
{Radhakrishnan}, V., \& {Cooke}, D.~J. 1969, Astrophys. Lett., 3, 225
%
\bibitem[\protect\citename{{Rankin}, }1986]{Rankin86}
{Rankin}, J.~M. 1986, ApJ, 301, 901
%
\bibitem[\protect\citename{{Rankin}, }1993a]{Rankin93a}
{Rankin}, J.~M. 1993a, ApJ, 405, 285
%
\bibitem[\protect\citename{{Rankin}, }1993b]{Rankin93b}
{Rankin}, J.~M. 1993b, ApJS, 85, 145
%
\bibitem[Rankin \& Ramachandran(2003)]{RR03}
Rankin, J.~M., \& Ramachandran, R.\ 2003, ApJ, 590, 411 
%
\bibitem[\protect\citename{{Rankin} {\em et~al.\ }\relax, }2003]{RSD03}
{Rankin}, J.~M., {Suleymanova}, S.~A., \& {Deshpande}, A.~A. 2003, MNRAS, 340, 1076
%
\bibitem[Rankin \& Wright(2007)]{RW07}
Rankin, J.~M., \& Wright, G.~A.~E.\ 2007, MNRAS, 379, 507
%
\bibitem[Rankin et al.(2013)]{RWB13}
Rankin, J.~M., Wright, G.~A.~E., \& Brown, A.~M.\ 2013, MNRAS, 433, 445
%
\bibitem[Redman et al.(2005)]{RWR05}
Redman, S.~L., Wright, G.~A.~E., \& Rankin, J.~M.\ 2005, MNRAS, 357, 859
%
\bibitem[\protect\citename{{Serylak} {\em et~al.\ }\relax, }2009]{SSW09}
{Serylak}, M., {Stappers}, B.~W., \& {Weltevrede}, P. 2009, A\&A, 506, 865
%
\bibitem[\protect\citename{{Smith} {\em et~al.\ }\relax, }2013]{SRM13}
{Smith}, E., {Rankin}, J., \& {Mitra}, D. 2013, MNRAS, 435, 1984
%
\bibitem[\protect\citename{{Srostlik} \& {Rankin}, }2005]{SR05}
{Srostlik}, Z., \& {Rankin}, J.~M. 2005, MNRAS, 362, 1121 (SR05)
%
\bibitem[\protect\citename{{Unwin} {\em et~al.\ }\relax, }1978]{Unwin78}
{Unwin}, S.~C., {Readhead}, A.~C.~S., {Wilkinson}, P.~N., \& {Ewing}, W.~S.\ 1978, MNRAS, 182, 711
%
\bibitem[\protect\citename{{van Leeuwen} {\em et~al.\ }\relax, }2002]{van02}
{van Leeuwen}, A.~G.~J., {Kouwenhoven}, M.~L.~A., {Ramachandran}, R., {Rankin},
  J.~M., \& {Stappers}, B.~W. 2002, A\&A, 387, 169
%
\end{thebibliography}
\end{document}